\RequirePackage{fix-cm}
\documentclass{svjour3}                     \smartqed

\usepackage{uclrn}
\usepackage{newtxtext}
\usepackage{newtxmath}
\usepackage[round]{natbib}

\usepackage{graphicx}
\usepackage{tabularx}
\usepackage{listings}
\usepackage{float}
\usepackage{multirow}         
\usepackage{mathtools}
\usepackage{makecell}
\usepackage{algorithm}
\usepackage{algpseudocode}
\usepackage{pifont}
\usepackage{verbatim}
\usepackage{enumitem}
\usepackage{url}
\usepackage{caption}
\usepackage{booktabs} \usepackage{subcaption}
\usepackage[misc,geometry]{ifsym} 
\usepackage{etoolbox}
\usepackage[colorlinks=true,allcolors=blue]{hyperref}

\rnnumber{17/10}

\begin{document}

\title{Awareness and Experience of Developers to\\ Outdated and License-Violating Code on Stack Overflow:\\ An Online Survey}

\titlerunning{Awareness and Experience of Developers to Outdated and License-Violating Code on Stack Overflow}        

\author{Chaiyong~Ragkhitwetsagul, Jens~Krinke, Rocco~Oliveto}

\institute{Chaiyong Ragkhitwetsagul~(\Letter), Jens Krinke \at
              Computer Science Department., University College London, UK \\
              Tel.: +44 (0)20 7679,  Fax: +44 (0)20 7387 1397\\
              \email{\{ucabagk, j.krinke\}@ucl.ac.uk}           }

\date{November 13, 2017}

\maketitle

\begin{abstract}
We performed two online surveys of Stack Overflow answerers and visitors to
assess their awareness to outdated code and software licenses in Stack Overflow
answerers. The answerer survey targeted 607 highly reputed Stack Overflow users
and received a high response rate of 33\%. Our findings are as follows.
Although most of the code snippets in the answers are written from scratch,
there are code snippets cloned from the corresponding questions, from personal or company projects,
or from open source projects. Stack Overflow answerers are aware that some of
their snippets are outdated. However, 19\% of the participants report that they
rarely or never fix their outdated code. At least 98\% of the answerers never include
software licenses in their snippets and 69\% never check for licensing conflicts
with Stack Overflow's CC BY-SA 3.0 if they copy the code from other sources to
Stack Overflow answers. The visitor survey uses convenient sampling and received
89 responses. We found that 66\% of the participants experienced a problem from
cloning and reusing Stack Overflow snippets. Fifty-six percent of the visitors
never reported the problems back to the Stack Overflow post. Eighty-five percent of the
participants are not aware that StackOverflow applies the CC BY-SA 3.0 license, and sixty-two percent never
give attributions to Stack Overflow posts or answers they copied the code from.
Moreover, 66\% of the participants do not check for licensing conflicts between
the copied Stack Overflow code and their software. With these findings, we
suggest Stack Overflow raise awareness of their users, both answerers and
visitors, to the problem of outdated and license-violating code snippets.
\end{abstract}

\section{Introduction}

Recent research shows that outdated third-party code and software licensing
conflicts are ramifications by code cloning, i.e.~reusing source code by copying
and pasting. \cite{Xia2014} report that a large number of open source systems
reuse outdated third-party libraries from popular open source projects. Using
the outdated code give detrimental effects to the software since they may
introduce vulnerabilities. On the other hand, \cite{German2009} found that code
cloning leads to software license conflicts among different systems.

The Internet encourages fast and easy code cloning by sharing and copying the
code to and from online sources. Developers nowadays do not only clone code from
local software projects, but also from online sources such as programming Q\&A
website and online code
repositories~\citep{Acar2016,Abdalkareem2017,An2017,Yang2017}. Stack Overflow is
one of the most popular programming Q\&A websites in the world. It has 7.6
million users, 14 million questions, 23 million answers\footnote{Data as of 21
	August 2017 from~\url{https://stackexchange.com/sites}}, and more than 50
million developers visiting each month\footnote{Data as of 21 August 2017 from:
	\url{https://stackoverflow.com}}. Some code snippets on Stack Overflow are found
to be problematic. \cite{Acar2016} discovered that many code snippets provided
as solutions on Stack Overflow are workarounds and occasionally contain defects
or vulnerabilities. They performed a user study and found that although Stack
Overflow helps developers to solve Android programming problems quicker than
other resources, at the same time, offers less secure code than books or the
official Android documentation. Only 17\% of the Stack Overflow discussion
threads the participant visited during the study contained secure code snippets.
They also found a similar piece of an API call copied from Stack Overflow by
participants in their study occurring in a random sample of 200,000 Android apps
from Google Play. In addition, \cite{An2017} investigated clones between 399
Android apps and Stack Overflow posts. They found 1,226 code snippets that were
reused from 68 Android apps. They also observed that there are 1,279 cases of
potential license violations.

Asking and answering questions on Stack Overflow involves including source code
snippets, either in a question or an answer or both. While many answers contain
snippets that are written from scratch, there are several answers containing
code snippets copied from other sources. The copied snippets on Stack Overflow
are snapshots of the code at the time of copying. They are less frequently
updated than in normal software projects and might not be up-to-date with their
originals, which are further modified due to bug fixing or feature improvements.
Besides, some snippets are copied from software systems with stricter licenses
than Stack Overflow's Attribution-ShareAlike 3.0 Unported (CC BY-SA 3.0). Prior
to this study, we performed a quantitative empirical study using code clone
detection tools to locate cloned code snippets in accepted answers. We found
that many snippets were copied from open source projects. Some cloned snippets
are outdated (or obsolete) and some possibly violate the original software
license. These snippets are potentially harmful to reuse. To accompany the
findings and gain insights into the cause of the problems, we resort to a
qualitative study using an online survey of Stack Overflow developers who 1)
regularly answer programming questions with code snippets and 2) reuse code
snippets from Stack Overflow. \textbf{The study aims to understand the
	developers' awareness and experience to outdated code and code licensing on
	Stack Overflow.}

In this paper, we use the term ``answerers'' to refer to Stack Overflow users
who actively answer questions, which is measured by their reputation. The
answerers gain a reputation from giving a helpful answer to a question and
receiving votes from other users. The reputation reflects trust they gain from
other users and also the quality of their answers. We ask the answerers the
origins of source code snippets in their answers, assess their awareness of
outdated and licensed code, and understand how they handle the issues.

We use the term ``visitors'' the refer to developers, who may or may not have a
Stack Overflow account but visit Stack Overflow when they encounter programming
problems. They copy code snippet(s) in a solution that is relevant to their
problem and reused them with or without a modification. We ask the visitors
about their experience of reusing source code from Stack Overflow and the
problems they faced including code obsolescence and software license.

We use the term ``license-violating code snippets'' or ``code with licensing conflicts''
interchangeably to refer to Stack Overflow cloned code snippets that \textit{potentially} violate the
original license by not including the original license statement in the cloned snippets and
adopt the Stack Overflow CC BY-SA 3.0 license instead.

\section{Research Methodology}
We followed the principles of survey research by~\cite{Pfleeger2001}
and~\cite{Kitchenham2002} by setting specific and measurable objective,
designing and scheduling the survey, selecting participants, analysing the data,
and reporting the results. We now discuss each of them in detail.

\subsection{Survey Objective}
The main objective of the survey is to understand the developers' awareness and
experience to outdated code and code licensing on Stack Overflow. It addresses
the following five research questions.

\begin{enumerate}
	\item \textbf{RQ1 (Sources of Stack Overflow Answer Snippets):} \textit{Where are the code snippets in Stack Overflow answers from?}
	\item \textbf{RQ2 (Awareness of Stack Overflow answerers to outdated code):} \textit{Are Stack Overflow answerers aware of outdated code in their answers?}
	\item \textbf{RQ3 (Awareness of Stack Overflow answerers to licensing violations):} \textit{Are Stack Overflow answerers aware of software licensing
		violations caused by code snippets in their answers?}
	\item \textbf{RQ4 (Problems from Stack Overflow code snippets):} \textit{What are the problems Stack Overflow visitors experiencing from reusing code snippets on Stack Overflow?}
	\item \textbf{RQ5 (Software license of Stack Overflow code
            snippets):} \textit{Are Stack Overflow visitors aware of code licensing on Stack Overflow?}
\end{enumerate}

The first three questions will be answered by
the Stack Overflow answerer survey, while the other two questions will be
answered by the Stack Overflow visitor survey.

\subsection{Survey Design and Schedule}
The study is conducted using unsupervised online surveys. We designed the surveys using
Google Forms. We created two versions of the
survey: \textbf{the answerer survey} and \textbf{the visitor survey}. Both
surveys are completely anonymous and the participants could decide to leave at
any time. They do not collect any sensitive personal information from the
participants and are approved for an ethical waiver by the designated ethics
officer in the Computer Science Department at University College London (UCL). 
The complete version of the surveys can be found in 
Appendix~\ref{appendixA}.

\subsubsection{The answerer survey} 
The survey contains 11 questions: 7 Likert's
scale questions, 3 yes/no questions, and one open-ended question for additional
comments. The first two questions are mandatory while the other 9 questions will
be shown to the participants based on their previous answers. The survey
collects information about the participant's software development experience, the
experience of answering Stack Overflow questions, sources of the Stack Overflow
snippets, awareness of outdated code in their answers, concerns regarding
license when they copy code snippets to Stack Overflow, and additional
feedback. The survey was open for participation for 50 days, from 25 July 2017
to 12 September 2017, before we collected and analysed the responses.

\subsubsection{The visitor survey} 
The survey consists of 16 questions: 9 Likert's
scale questions, 3 yes/no questions, 2 multiple-choice questions, and 2
open-ended questions. The first four questions are mandatory while the other 12
questions will be shown to the participants based on their previous answers. The
survey collects information about the participant's software development
experience, the importance of Stack Overflow, reasons for reusing Stack Overflow
snippets, problems from Stack Overflow snippets, licensing of code on Stack
Overflow, and additional feedback. The survey was open for participation for 2 months, 
from 25 July 2017 to 25 September 2017, before we collected and analysed
the responses.

\subsection{Participant Selection}

\subsubsection{The answerer survey} 
We selected the participants for our answerer survey based on their Stack
Overflow reputations. On Stack Overflow, a user's reputation reflects how much
the community trusts the user. A user earns reputation when he or she receives
upvotes for good questions and useful answers. For example, they gain reputation when they receive an upvote for
their question (+5) or their answer (+10), or when their answer is accepted (+15)\footnote{Stack
	Overflow Reputation:~\url{https://stackoverflow.com/help/whats-reputation}}. Thus, Stack
Overflow reputation is an indicator of user's skills and their involvement in
asking and answering questions on the site.

\begin{table}
	\centering
	\caption{The Stack Overflow answerer taken the surveys}
	\label{tab:answerers}
	\begin{tabular}{lrrrr}
		\toprule
		Target group & Reputation & Sent emails & Answers & Rate \\
		\midrule
		Answerer Group 1 & 963,731--7,674 & 300 & 117 & 39\% \\
		Answerer Group 2 & 7,636--6,999 & 307 & 84 & 27\% \\
		\midrule
		Total & -- & 607 & 201 & 33\% \\
		\bottomrule
	\end{tabular}
\end{table}

The participants were invited to take the survey via email addresses publicly
available on their Stack Overflow and GitHub profiles. We selected the answerers
based on their all-time reputation ranking\footnote{Stack Overflow Users (data as of
	25 July 2017):~\url{https://stackoverflow.com/users?tab=Reputation&filter=all}}
and separated them into two groups (see Table~\ref{tab:answerers}) so that we
can compare the results or observe similar patterns between them. The first
group had a reputation from 963,731 (the highest) to 7,674, and the second group
had a reputation from 7,636 to 6,999. We sent out 300 and 307 emails
(excluding undelivered ones) to the two groups respectively. 

\subsubsection{The visitor survey} 
We adopted non-probability convenient sampling to invite participants for this
survey. Participation in the survey requires experience of visiting Stack
Overflow for solving programming tasks at least once. The participants were
invited to take the survey via five channels. The first channel is via the first
author's social media post (Facebook) inviting software developers who have
experience of copying code snippets from Stack Overflow to take the survey. The
second channel is a popular technology news and media community in Thailand
called \textsf{blognone.com} which attracts a high number of Thai software
developers. The first author posted an invitation to the visitor survey in a
discussion forum mentioning the requirements to take the survey. The third
channel collected answers from the University of Molise in Italy, where the third author
works. The last two channels are the
\textsf{comp.lang.java.programmer} group and the Software Engineering
Facebook page. The number of participants taken the survey is shown in
Table~\ref{tab:visitors}.

\begin{table}
	\centering
	\caption{The Stack Overflow visitors taken the survey}
	\label{tab:visitors}
	\begin{tabular}{lr}
		\toprule
		Group & Answers \\
		\midrule
		Social media (Facebook posts) & 47 \\
		Blognone.com & 32 \\
		University of Molise & 6 \\
		comp.lang.java.programmer & 3 \\
		Software Engineering Facebook page & 1 \\
		\midrule
		Total & 89 \\
		\bottomrule
	\end{tabular}
\end{table}

\subsection{Data Analysis}
Google Forms provide a helpful summary of responses to the online surveys. We
relied on the summary from Google Forms when analysing the answers from the
answerer survey. For the visitor survey, since we created a copy of the survey for each
group of the participants, we downloaded the responses in comma-separated values
(CSV) files and merged the results before the analysis.

\section{Results and Discussions}
We collected and analysed the results after we closed the surveys on 12
September 2017. We now discuss the results from the answerer and
visitor survey separately.

\subsection{The answerer survey}
We received 117 answers (39\% response rate) from the first group and 84 answers
(27\% response rate) from the second group of Stack Overflow answerers. The response
rate from both groups was high considering other online surveys in software
engineering~\citep{Punter2003}.

\subsubsection{General Information}

\begin{figure}
	\begin{subfigure}{.5\textwidth}
		\centering
		\includegraphics[width=0.8\linewidth]{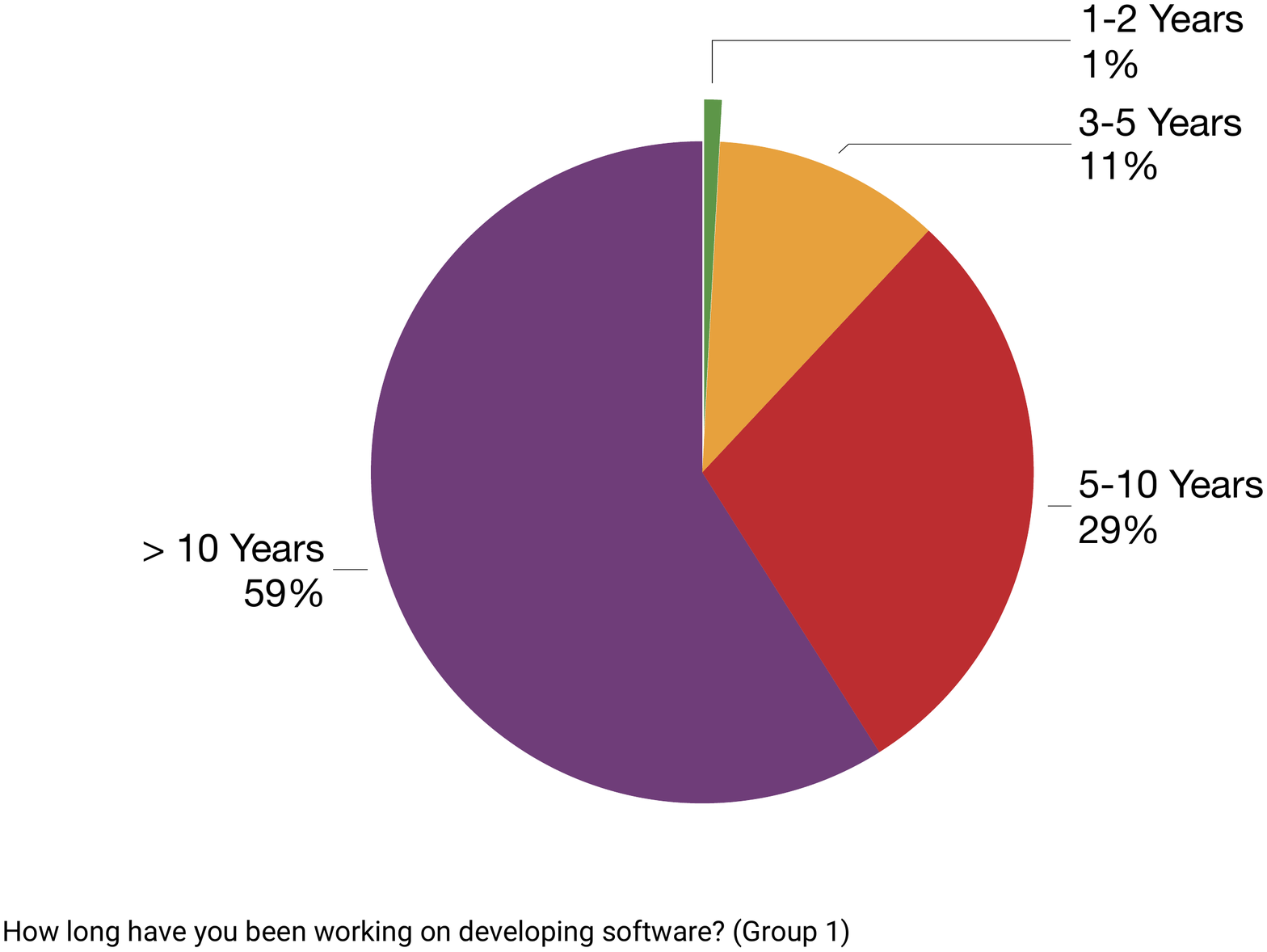}
		\caption{Group 1}
		\label{fig:survey_group1_exp}
	\end{subfigure}	\begin{subfigure}{.5\textwidth}
		\centering
		\includegraphics[width=0.8\linewidth]{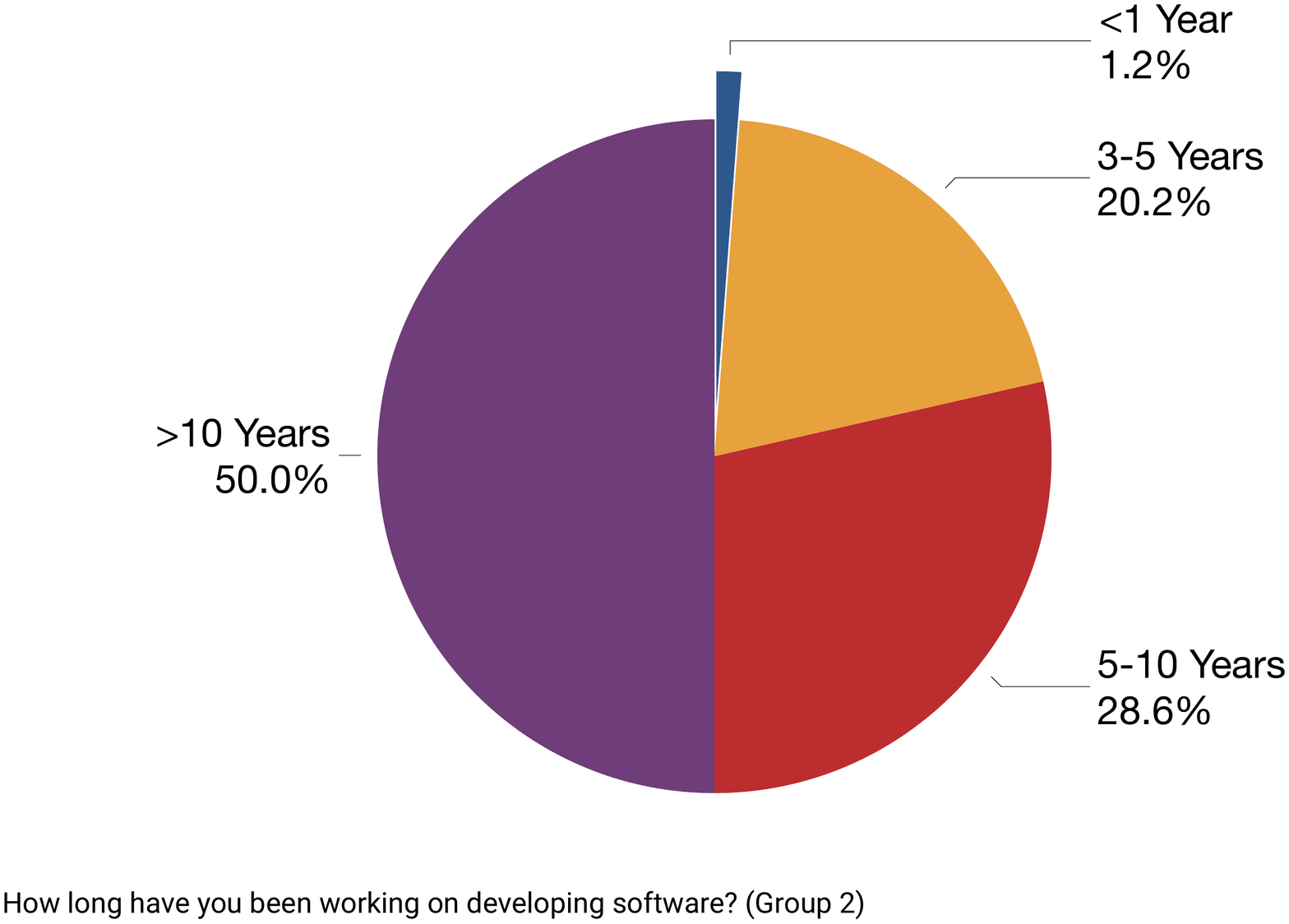}
		\caption{Group 2}
		\label{fig:survey_group2_exp}
	\end{subfigure}
	\caption{Experience of the Stack Overflow answerers}
	\label{fig:survey_exp}
\end{figure}

The majority of users in both groups are experienced developers with more than
10 years of experience or between 5 to 10 years as depicted in
Figure~\ref{fig:survey_exp}. There are 59\% of the answerers in Group 1 
and 50\% of the answerers in Group 2 that
have more than 10 years of software development experience.

\begin{figure}
		\centering
		\includegraphics[width=0.8\linewidth]{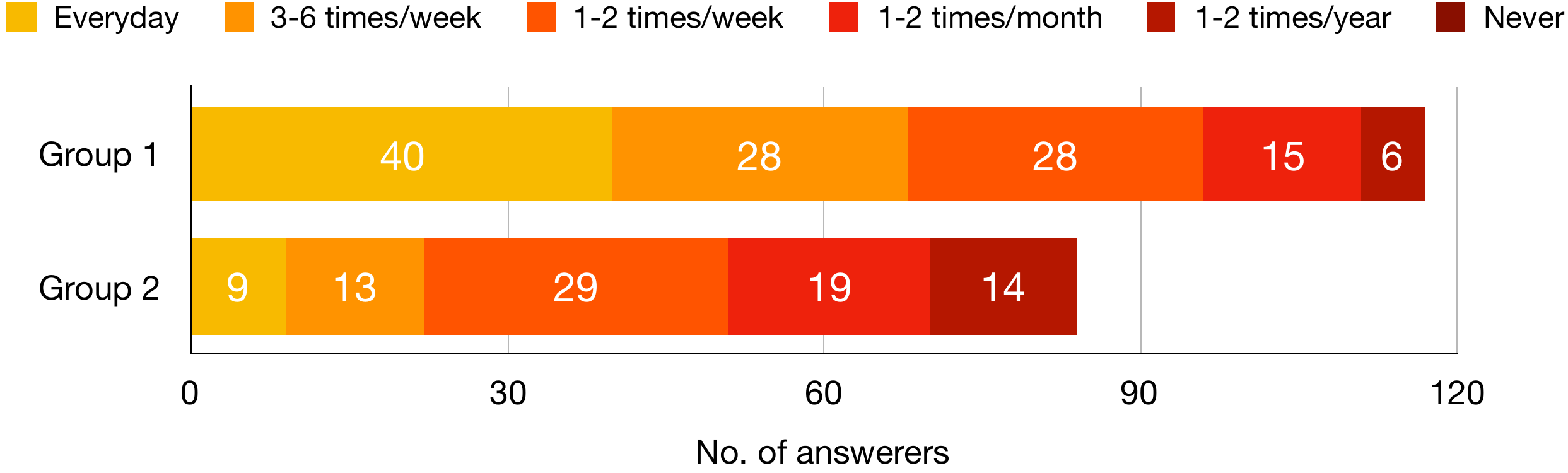}
		\caption{Frequency of answering questions}
		\label{fig:survey_answer_frequency}
\end{figure}

\begin{figure}
	\begin{subfigure}{.5\textwidth}
		\centering
		\includegraphics[width=\linewidth]{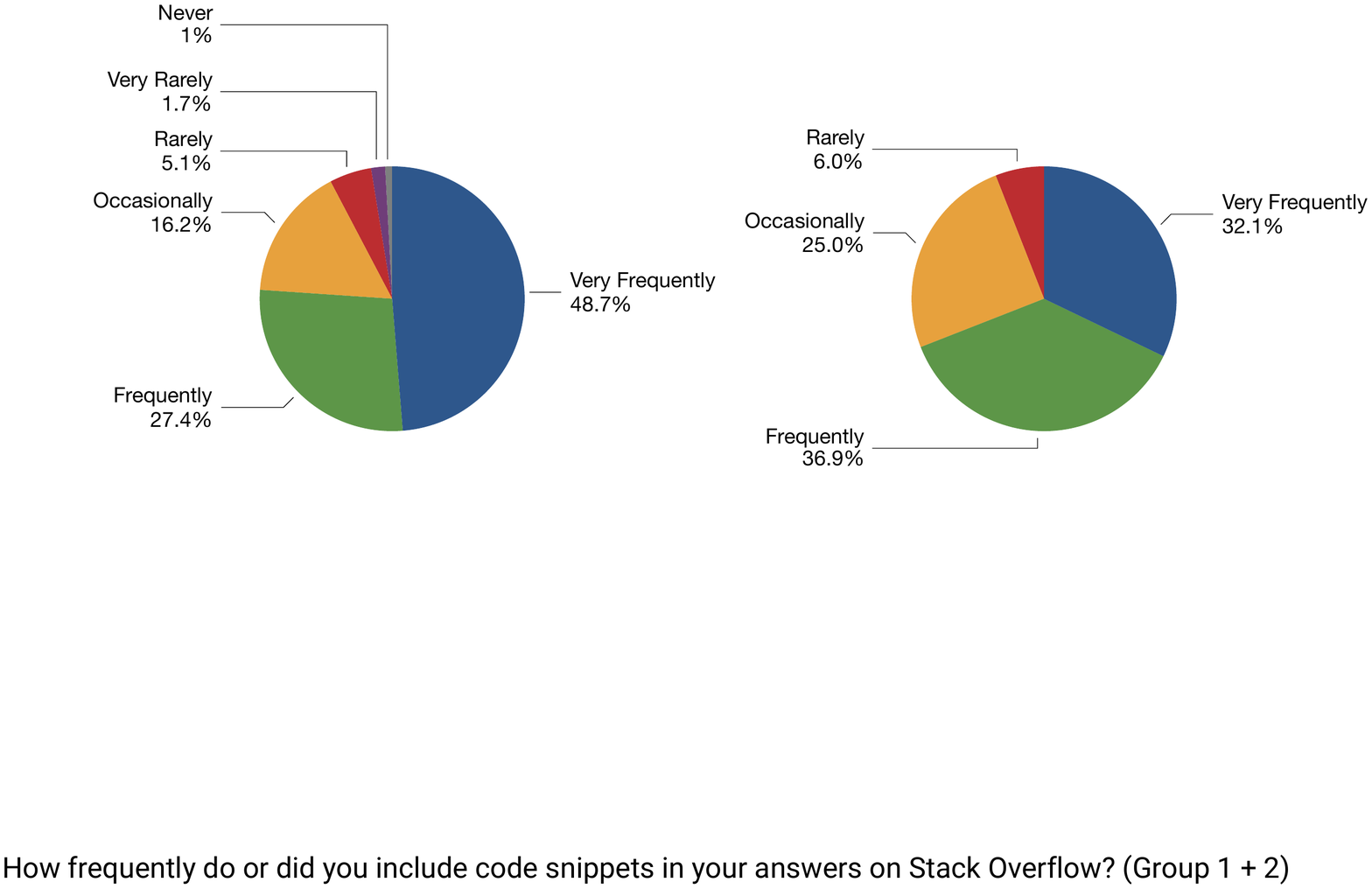}
		\caption{Group 1}
		\label{fig:survey_answer_frequency_with_code_1}
	\end{subfigure}	\begin{subfigure}{.5\textwidth}
		\centering
		\includegraphics[width=\linewidth]{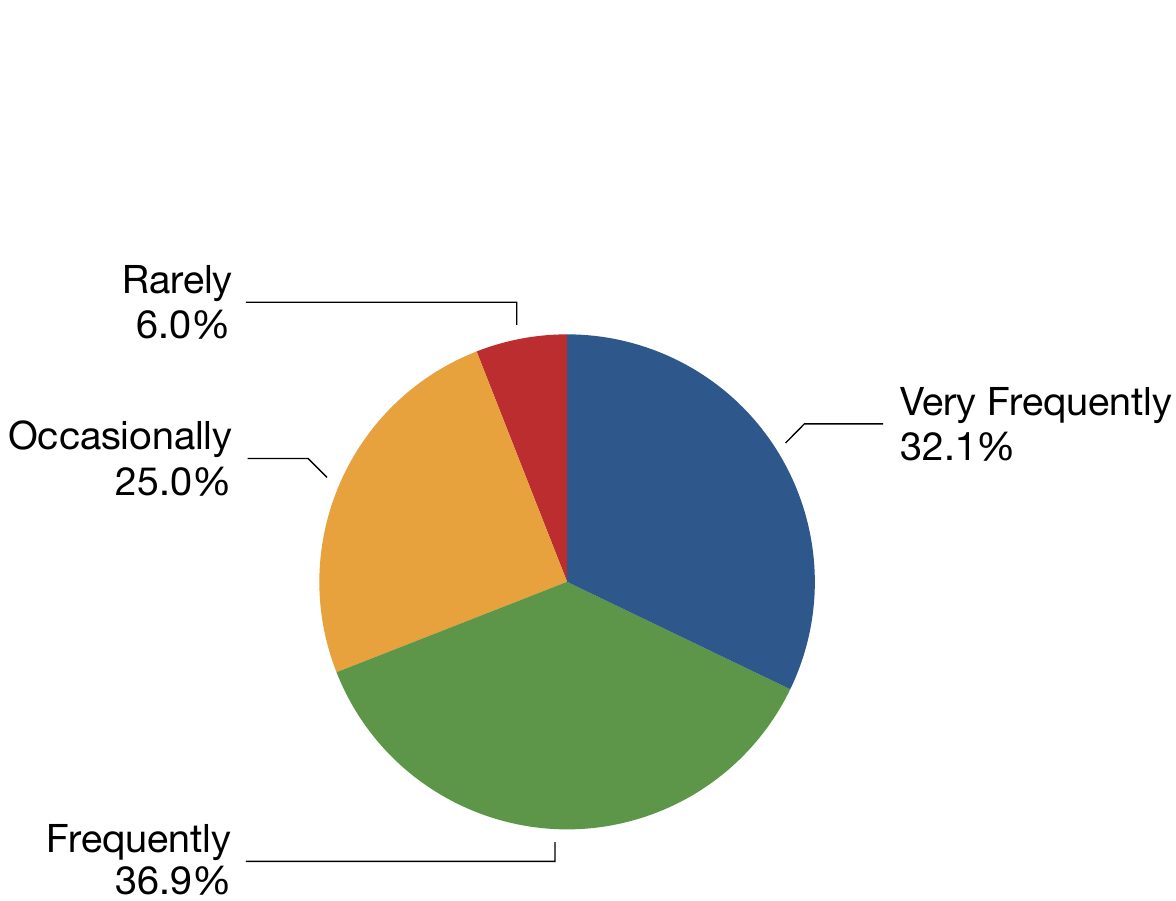}
		\caption{Group 2}
		\label{fig:survey_answer_frequency_with_code_2}
	\end{subfigure}
	\caption{Frequency of answering questions with code snippet(s)}
	\label{fig:survey_answer_frequency_with_code}
\end{figure}

The participants are active users and regularly answer questions on Stack
Overflow (see Figure~\ref{fig:survey_answer_frequency}). Eighty-two (82\%) and
sixty (61\%) of answerers from Group 1 and Group 2  answer questions at least
once a week. More than half of the answerers very frequently (81--100\% of the
time) or frequently (61--80\% of the time) include code snippets in their
answers. To break down into two groups as depicted in
Figure~\ref{fig:survey_answer_frequency_with_code}, Group 1 very frequently
(48.7\%) and frequently (27.4\%) provide code examples when answering. Likewise,
Group 2 follow the same trend (very frequently for 32.1\% and frequently for
36.9\%). Interestingly, there is one participant in the first group who never includes code snippet
in his/her answer thus the results after this question is from 116
participants of the first group.

\subsubsection*{RQ1: Where are the code snippets in Stack Overflow answers from?} 

\begin{figure*}
	\begin{subfigure}{.5\textwidth}
		\centering
		\includegraphics[width=.9\linewidth]{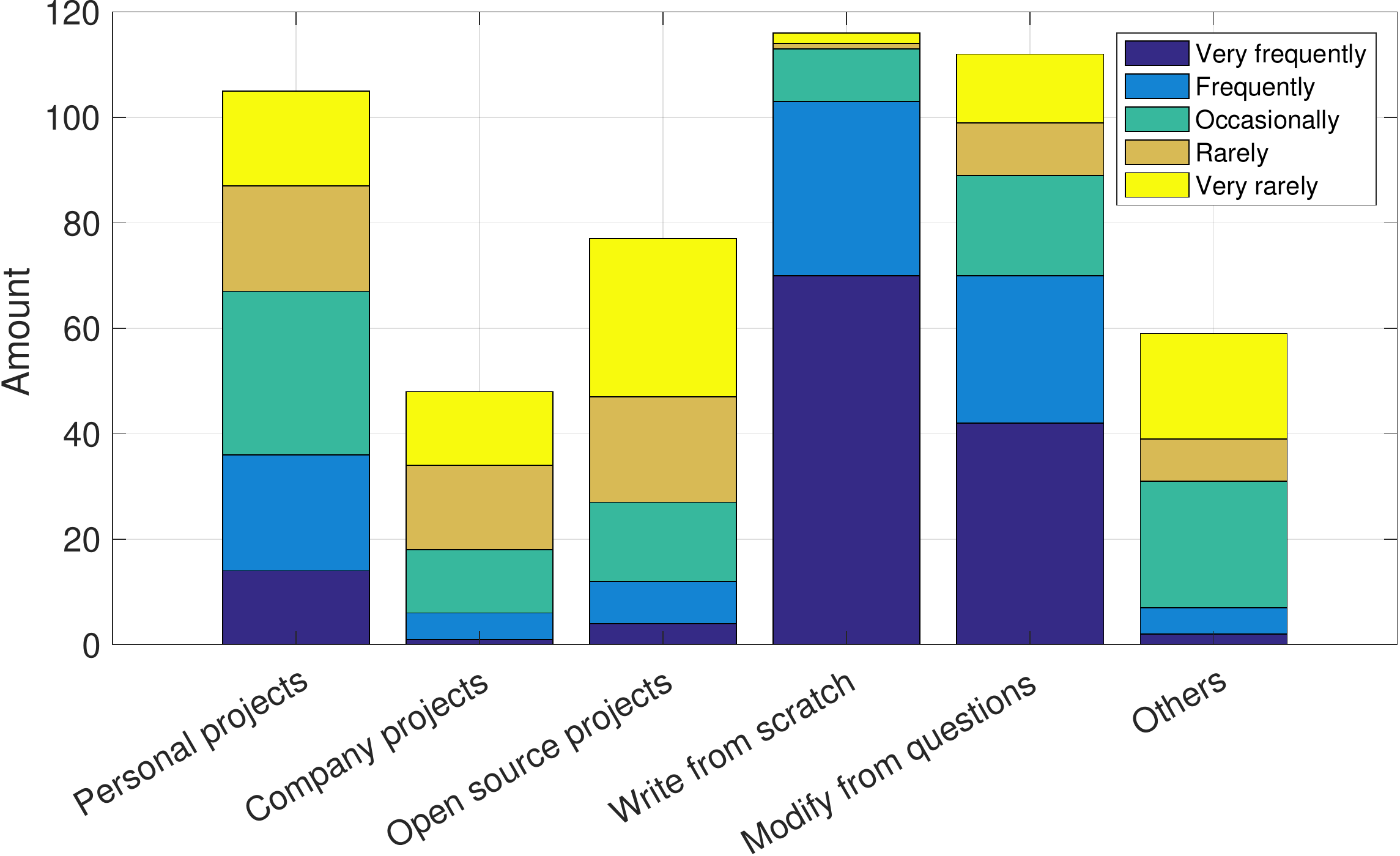}
		\caption{Group 1}
		\label{fig:survey_snippet_source_1}
	\end{subfigure}	\begin{subfigure}{.5\textwidth}
		\centering
		\includegraphics[width=.9\linewidth]{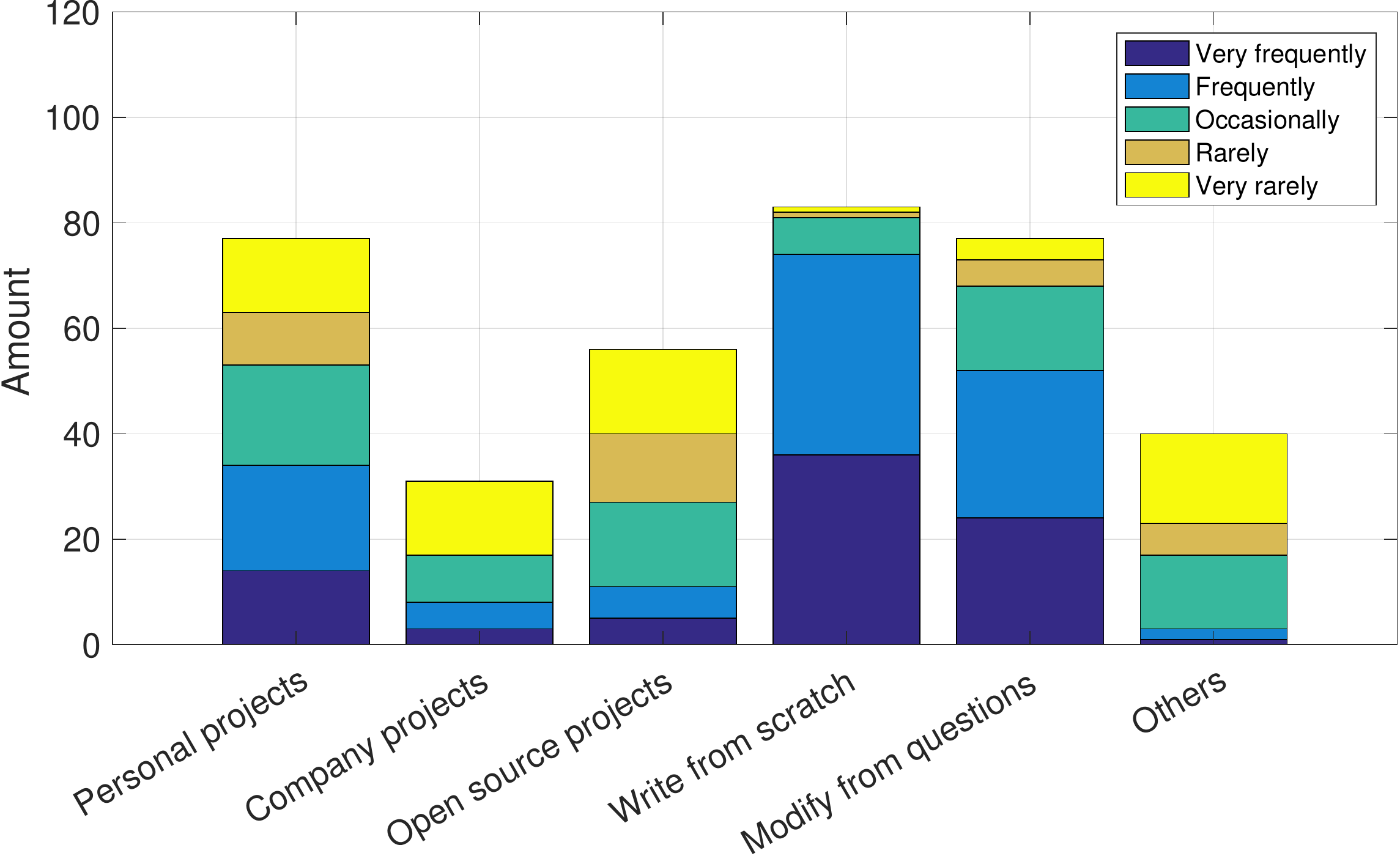}
		\caption{Group 2}
		\label{fig:survey_snippet_source_2}
	\end{subfigure}
	\caption{Sources of code snippets in Stack Overflow answers}
	\label{fig:survey_snippet_source}
\end{figure*}

To answer this research question, we asked the participants for the original
source of their code examples. We provided six options (allowing more than one
answer) including \textit{I copied them from my own personal projects},
\textit{I copied them from my company's projects}, \textit{I copied them from
	open source projects}, \textit{I wrote the new code from scratch}, \textit{I
	copied the code from the question and modify it for the answer}, and
\textit{Others}. The answers are shown in
Figure~\ref{fig:survey_snippet_source}. Participants in Group 1 mainly write new
code from scratch (116) or copy from the code snippets in question and modify it
for the answer (112), followed by from their personal projects (105), open
source projects (77), other sources (59), and company projects (48). For Group
2, the main source is also writing code from scratch (83), followed by copying
from personal projects and modifying from the question (77), copying from open
source projects (56), copying from other sources (40), and copying from company
projects (31). There are 133 answerers out of total 607 from two groups who have cloned
code snippets from open source projects into their answers at least once. We are interested
in this type of clones and will investigate further in the later RQs.

\vspace{0.5cm} \noindent\fbox{	\parbox[c][2cm]{0.98\textwidth}{        \textit{For RQ 1, we found that answering questions by writing the new
	code from scratch is the most popular choice for Stack Overflow
	answerers followed by modifying the code in question or copying from
	personal projects. Other less popular choices include copying code from
	open source projects and from other sources. Copying code from company
	projects are the least popular choice.} }}
\vspace{0.5cm}

\subsubsection*{RQ2: Are Stack Overflow answerers aware of outdated code in their answers?}

\begin{figure}
	\centering
	\includegraphics[width=.5\linewidth]{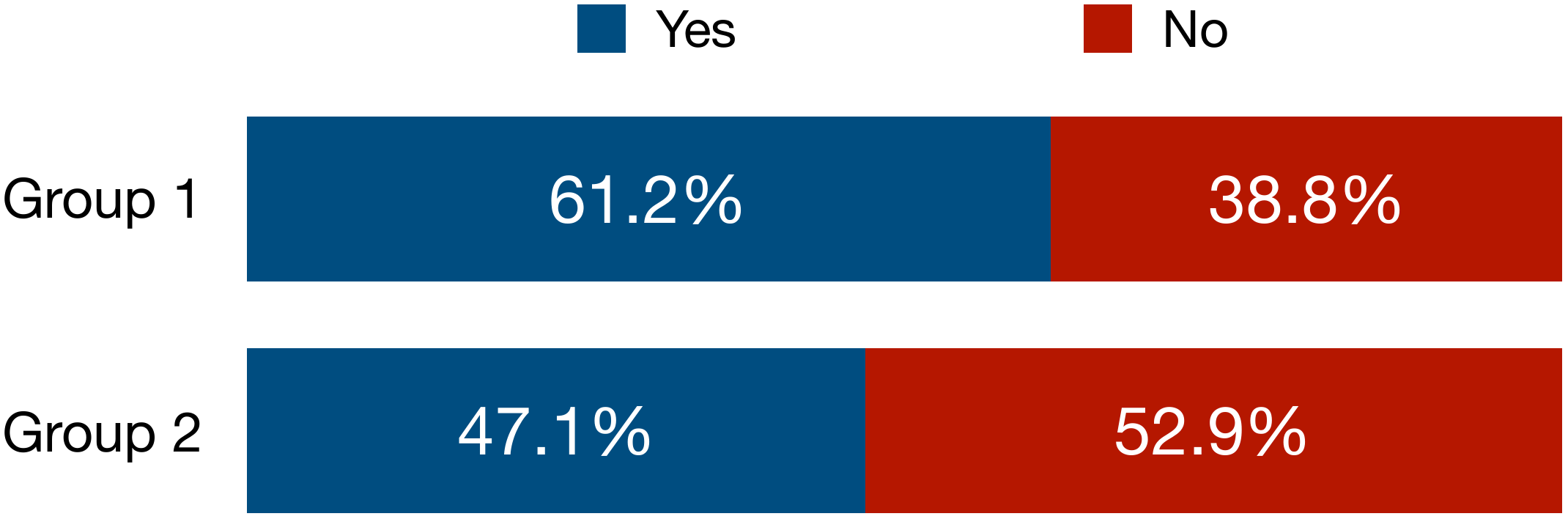}
	\caption{Percentage of answerers who are notified of outdated code in their Stack Overflow answers.}
	\label{fig:survey_outdated}
\end{figure}

\begin{figure}
	\begin{subfigure}{.5\textwidth}
		\centering
		\includegraphics[width=.8\linewidth]{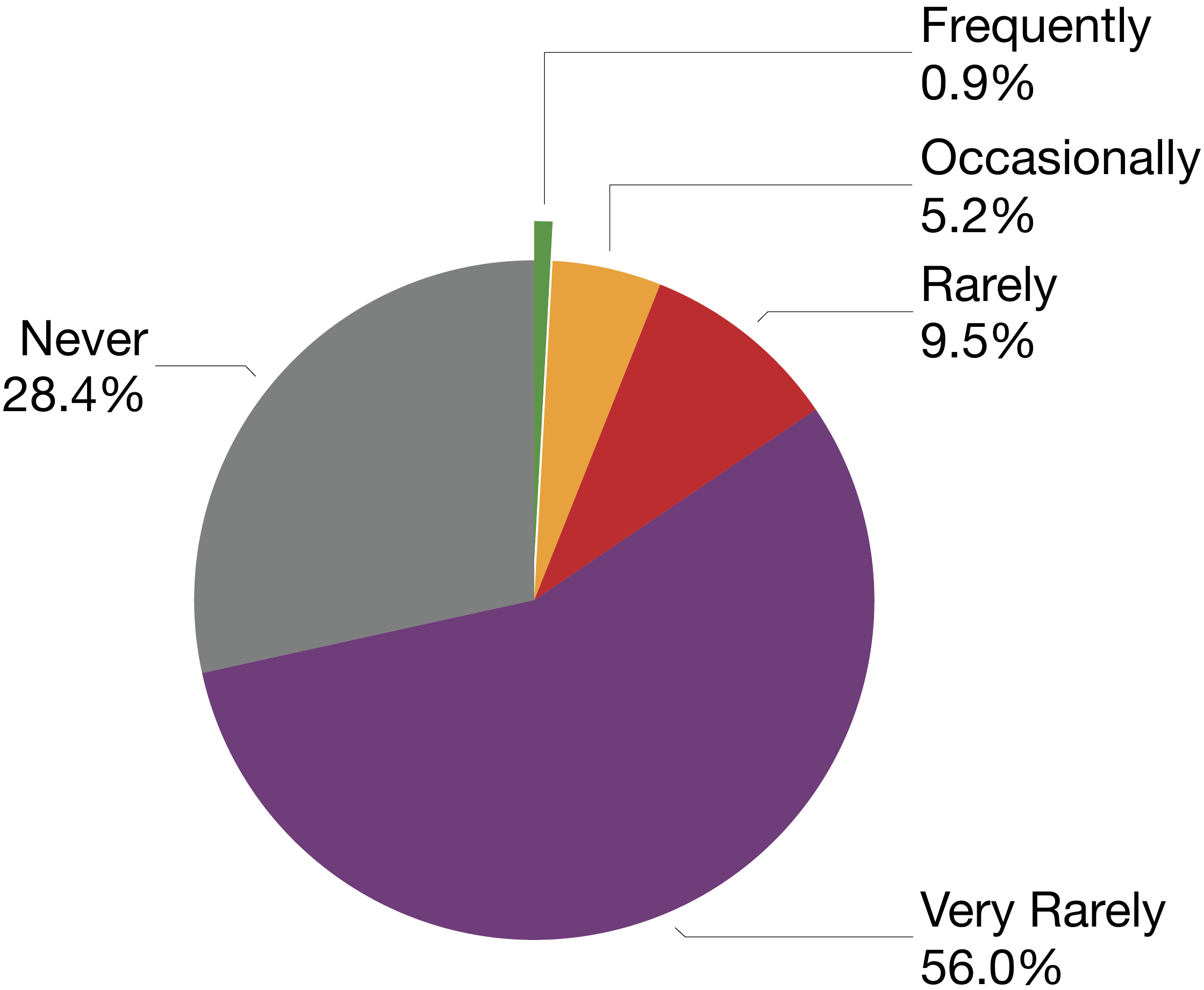}
		\caption{Group 1}
		\label{fig:survey_outdated_freq_1}
	\end{subfigure}	\begin{subfigure}{.5\textwidth}
		\centering
		\includegraphics[width=.8\linewidth]{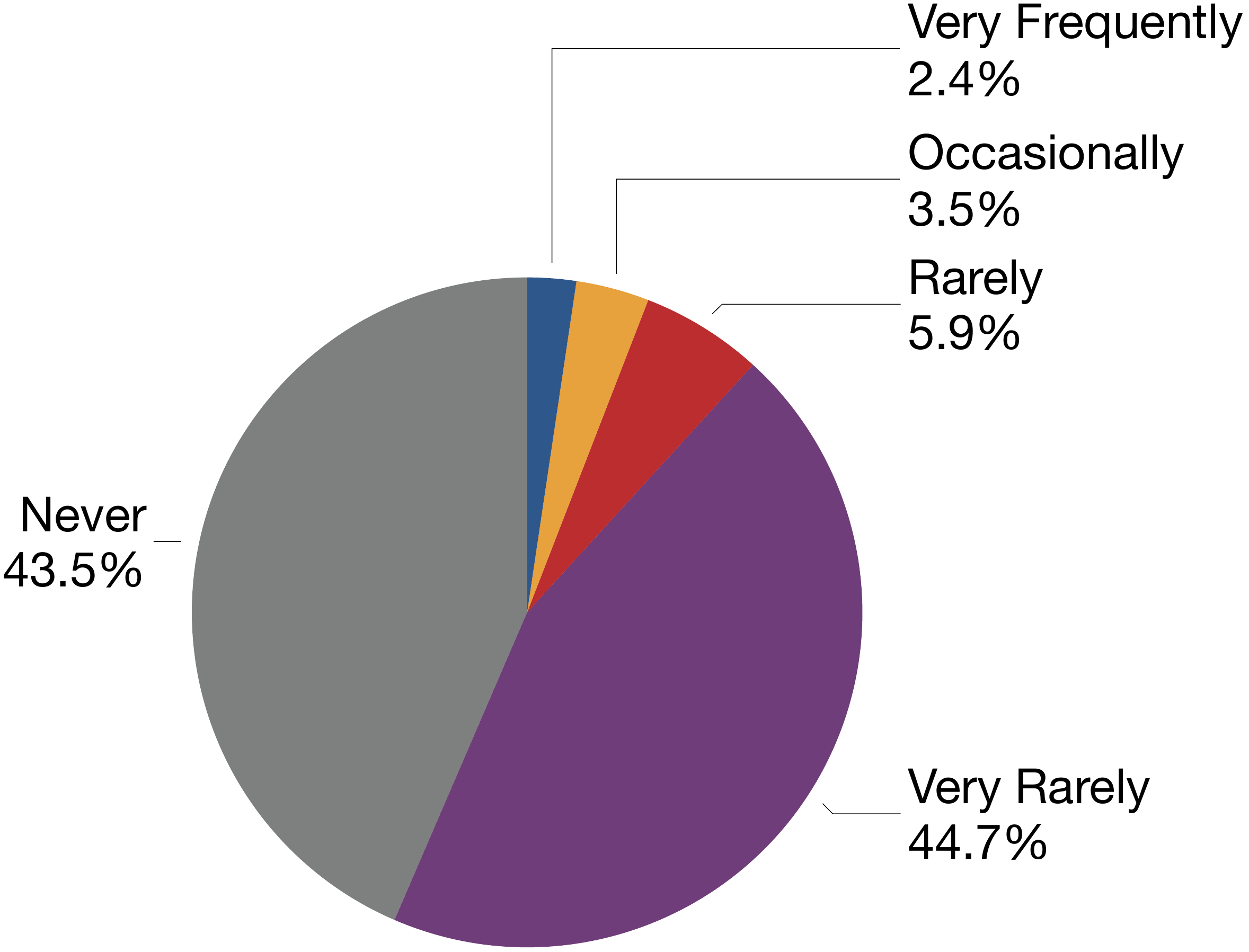}
		\caption{Group 2}
		\label{fig:survey_outdated_freq_2}
	\end{subfigure}
	\caption{Frequency of being notified of outdated code in the answerers' answers.}
	\label{fig:survey_outdated_freq}
\end{figure}

Half of the top answerers on Stack Overflow are aware of outdated code in their
answers. Seventy-one participants (61.2\%) of Stack Overflow answerers in Group
1 have been notified of outdated code in at least one of their answers. The
ratio drops to forty participants (47.1\%) in Group 2. We asked a follow up
question regarding the frequency of being notified of outdated code in their
answers. We found that only 0.5\% and 5.2\% of the answerers in Group 1
are frequently or occasionally been notified. The answerers in Group 2 have very
frequently and occasionally been notified for 2.4\% and 3.5\% respectively.
Please note that we found inconsistencies between the answers of these two
questions. The percentage of participants who have ``Never'' been notified of
outdated code in their Stack Overflow answers are 38.8\% and 52.9\% for Group 1
and Group 2 respectively. However, the answers for the frequency of being notified
equal to ``Never'' decrease to 28.4\% and 43.5\% for Group 1 and 2 respectively
(see
Figure~\ref{fig:survey_outdated}~and~Figure~\ref{fig:survey_outdated_freq}).

\begin{figure}
	\begin{subfigure}{.5\textwidth}
		\centering
		\includegraphics[width=.8\linewidth]{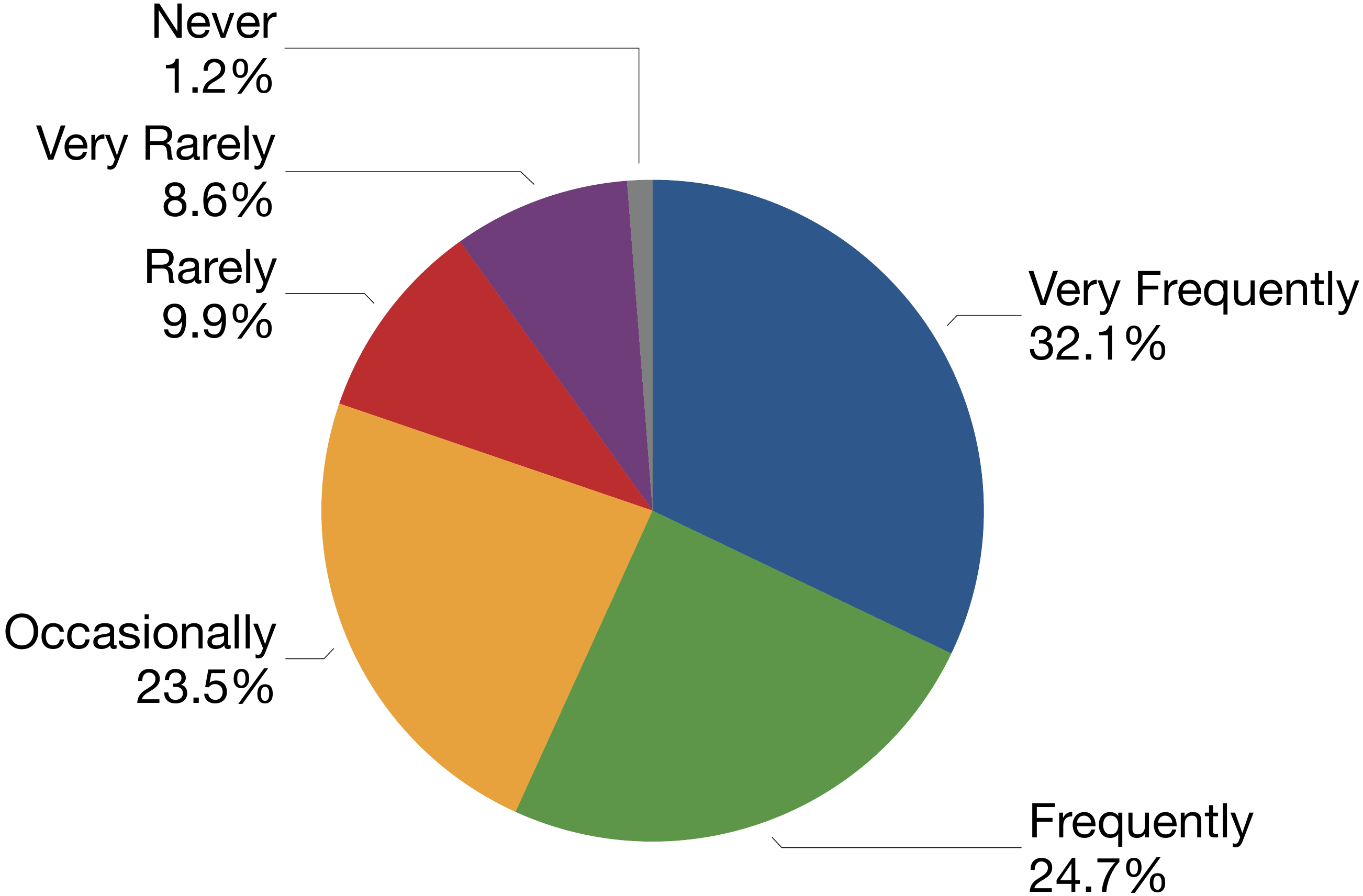}
		\caption{Group 1}
		\label{fig:survey_outdated_fix_1}
	\end{subfigure}	\begin{subfigure}{.5\textwidth}
		\centering
		\includegraphics[width=.8\linewidth]{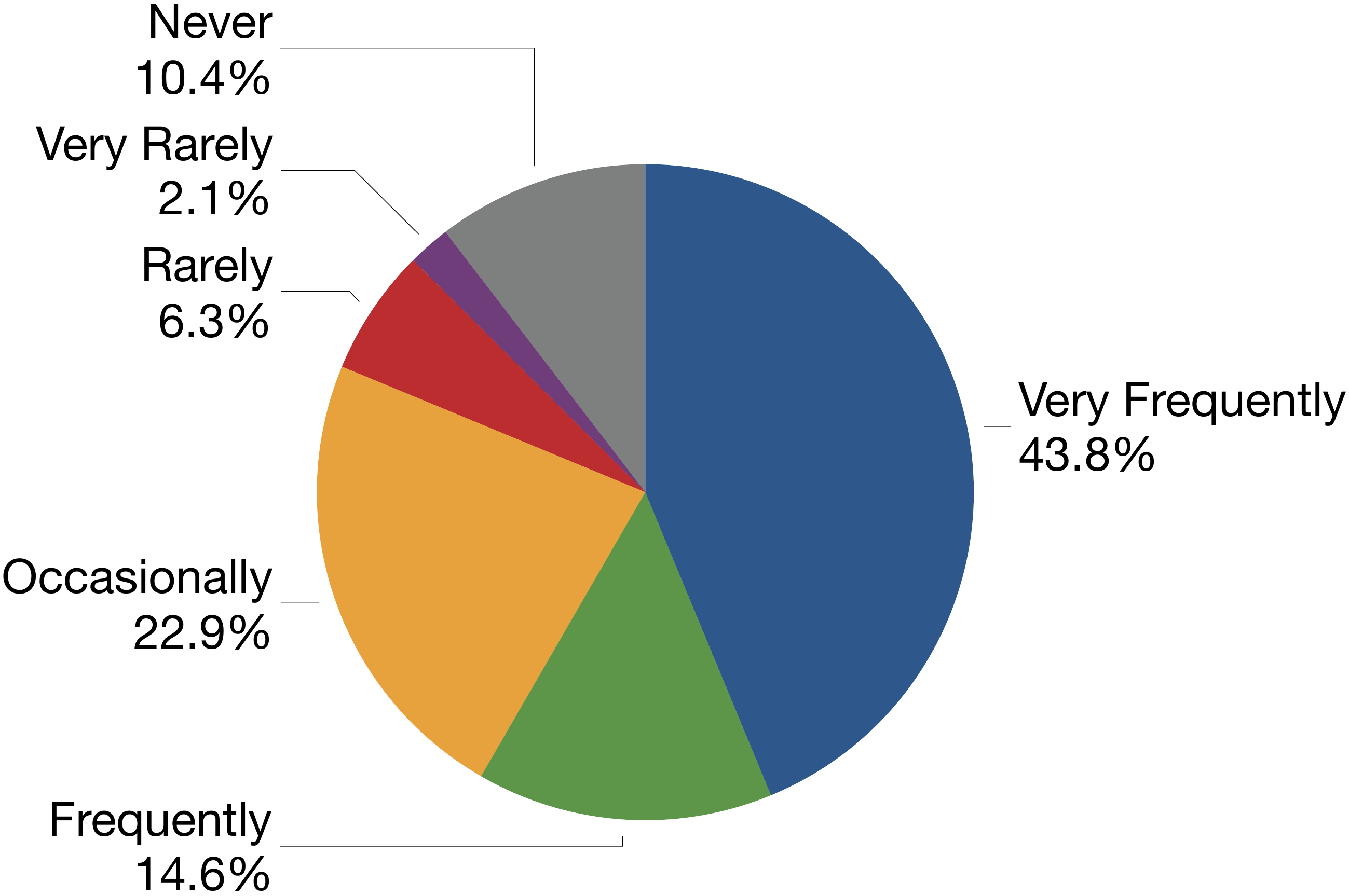}
		\caption{Group 2}
		\label{fig:survey_outdated_fix_2}
	\end{subfigure}
	\caption{Frequency of the answerers fixing their outdated code.}
	\label{fig:survey_outdated_fix}
\end{figure}

We then asked the participants who have been notified of their outdated code (83
and 48 participants from Group 1 and 2 respectively) a follow-up questions
\textit{``how frequently did you fix your outdated code on Stack Overflow?''}
The answers, depicted in Figure~\ref{fig:survey_outdated_fix}, show that more
than half of them frequently fix the outdated code snippets. However, there are
17 (19.8\%) and 9 (18.8\%) participants in Groups 1 and 2 who rarely, very
rarely or never fix their code.

Regarding the issue of outdated code, one participant expresses their concern
in the open comment question: \textit{``The main problem for me/us is outdated
	code, esp. as old answers have high Google rank so that is what people see
	first, then try and fail. Thats why we're moving more and more of those examples
	to knowledge base and docs and rather link to those.''} On the other hand,
another participant does not  worry about his/her outdated code and how he
or she handles them: \textit{``On
	the matter of deprecation, I almost entirely use .NET which has got different
	versions of the framework. Therefore, code deprecation is not often a problem
	since what is deprecated on one version of the framework may be the only way of
	solving a given problem on an older version of the framework. I may also have to
	add that questions I tend to answer are about how to solve general coding
	problems so they are not usually subject to deprecation.''}

\vspace{0.5cm} \noindent\fbox{	\parbox[c][1cm]{0.98\textwidth}{		\textit{For RQ 2, Stack Overflow answerers are aware of outdated code in their
			answers. Nonetheless, there are approximately 19\% of the answerers who rarely
			or never fix their outdated code for which
                        they have been notified.} }}
\vspace{0.5cm}

\subsubsection*{RQ3: Are Stack Overflow answerers aware of software licensing
	violations caused by code snippets in their answers?}

\begin{figure}
	\begin{subfigure}{.5\textwidth}
		\centering
		\includegraphics[width=.4\linewidth]{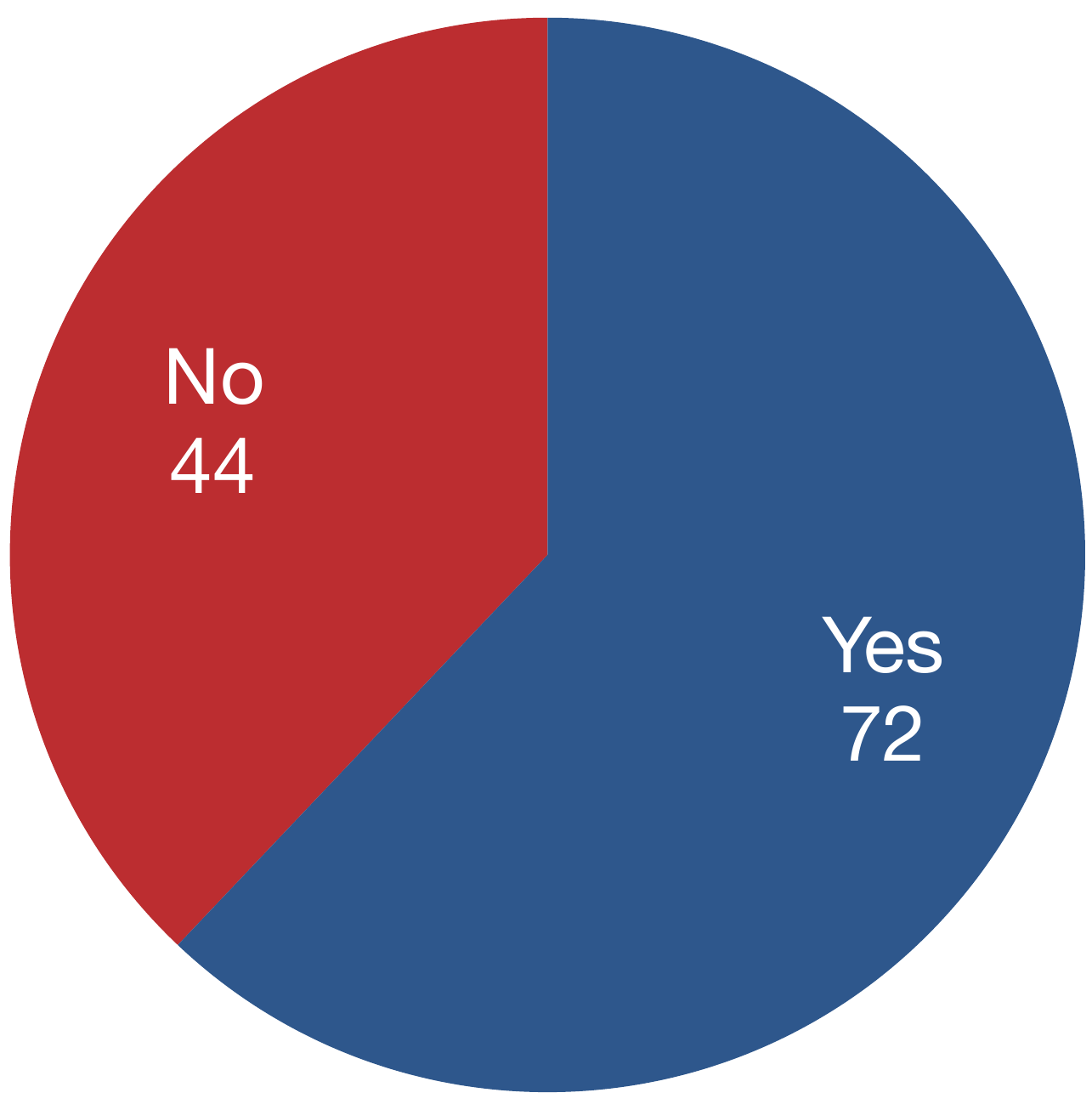}
		\caption{Group 1}
		\label{fig:survey_license_known_1}
	\end{subfigure}	\begin{subfigure}{.5\textwidth}
		\centering
		\includegraphics[width=.4\linewidth]{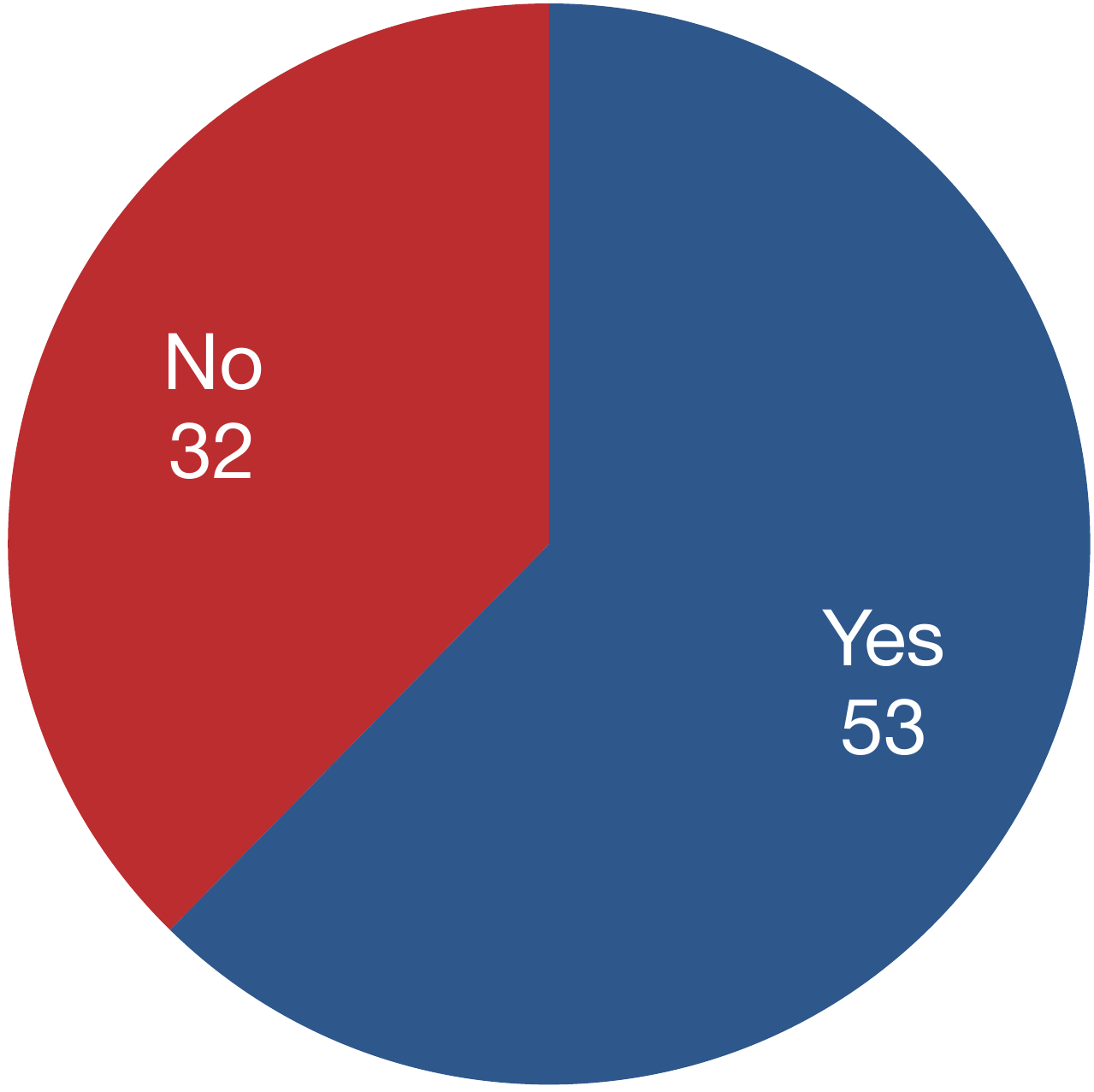}
		\caption{Group 2}
		\label{fig:survey_license_known_2}
	\end{subfigure}
	\caption{Awareness of answerers to Stack Overflow CC BY-SA 3.0 license}
	\label{fig:survey_license_known}
\end{figure}

As shown in Figure~\ref{fig:survey_license_known}, more than half of the
answerers in both groups, 72 (62.1\%) and 53 (62.3\%) respectively, are aware that Stack
Overflow apply Creative Commons Attribution-ShareAlike 3.0 Unported (CC BY-SA
3.0) to content in the posts, including code snippets, while the rest of 44 (37.9\%)
and 32 (37.6\%) are not. 

\begin{figure}
	\begin{subfigure}{.5\textwidth}
		\centering
		\includegraphics[width=.4\linewidth]{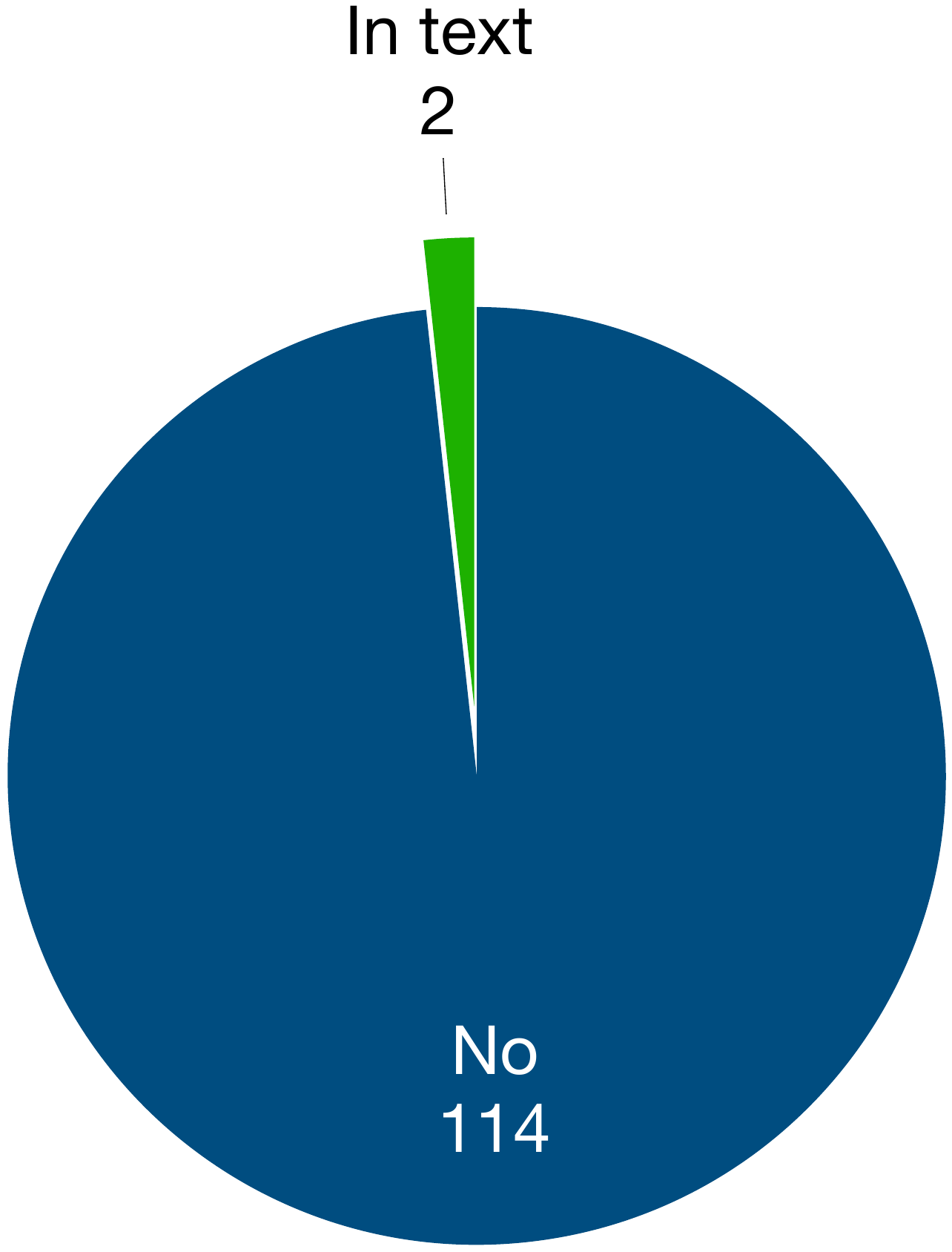}
		\caption{Group 1}
		\label{fig:survey_license_1}
	\end{subfigure}	\begin{subfigure}{.5\textwidth}
		\centering
		\includegraphics[width=.4\linewidth]{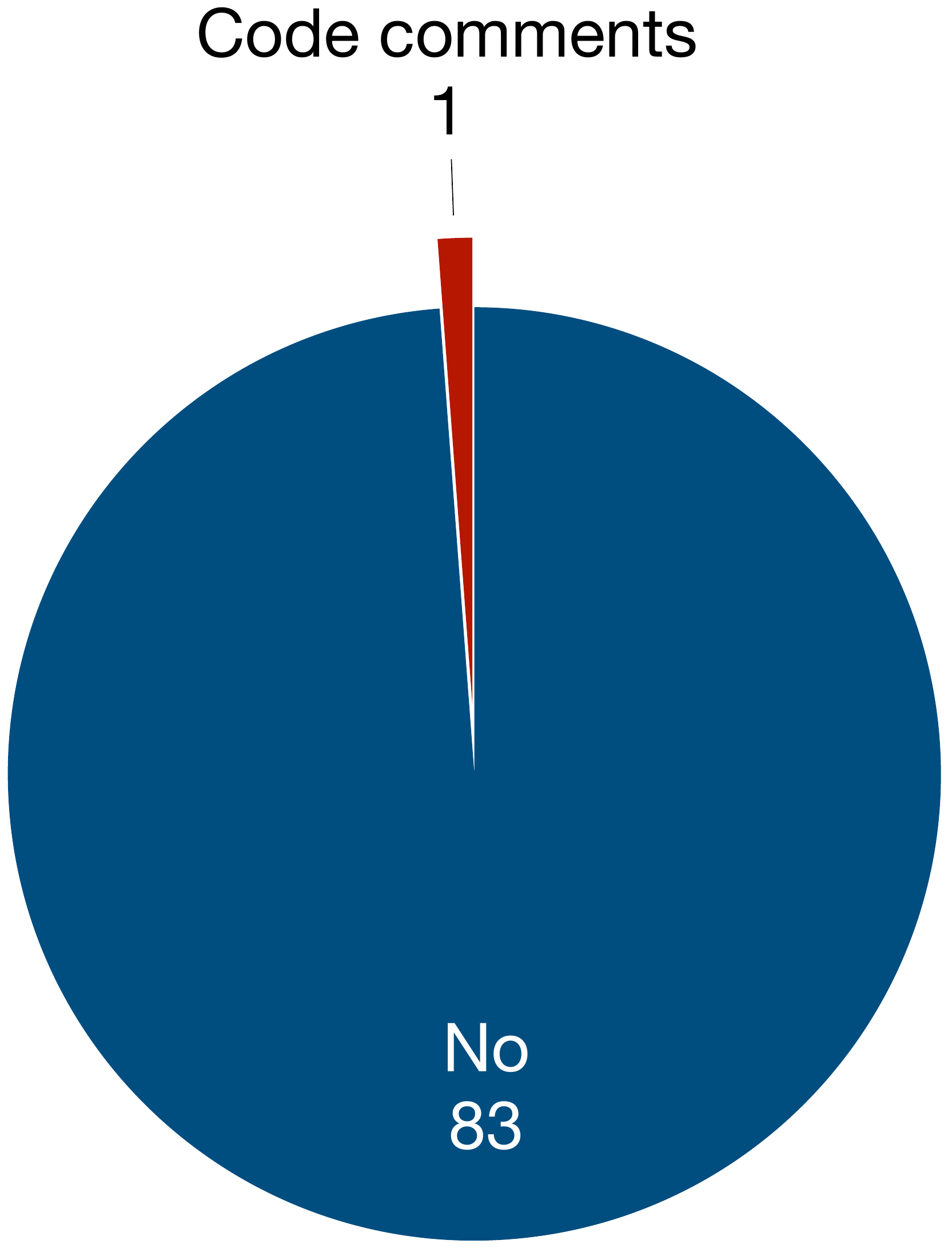}
		\caption{Group 2}
		\label{fig:survey_license_2}
	\end{subfigure}
	\caption{Software license in Stack Overflow code snippets.}
	\label{fig:survey_license}
\end{figure}

Almost every answerer in both groups, 114 out of 116 (98\%) and 84 out of 85
(99\%) respectively, do not include license statement in their code snippets
(numbers are shown in Figure~\ref{fig:survey_license}). Some of the participants
explain the reason in the open comment question which we summarised into three
groups as follows. First, they choose to post only their own code or code that
is adapted from the question. The code is automatically subjected to Stack
Overflow's CC BY-SA 3.0 without any explicit licensing statement. Second, they
copy the code from company or open source projects that they know are permitted
to be publicly distributed. Hence, no license statement is required. Third, some
answerers believe that code snippets in their answers are too small to claim any
intellectual property on them and fall under fair use~\citep{fairuse}.

While almost nobody explicitly includes a software license in their snippets,
many participants include a statement on their profile page that all their
answers are under a certain license. For example, \textit{All code posted by me on
	Stack Overflow should be considered public domain without copyright. For
	countries where public domain is not applicable, I hereby grant everyone the
	right to modify, use and redistribute any code posted by me on Stack Overflow
	for any purpose. It is provided "as-is" without warranty of any kind.}  Many
participants either declare their snippets to be public domain, or they grant
additional licenses, e.g.\ Apache 2.0 or MIT/Expat.

\begin{figure}
	\begin{subfigure}{.5\textwidth}
		\centering
		\includegraphics[width=.75\linewidth]{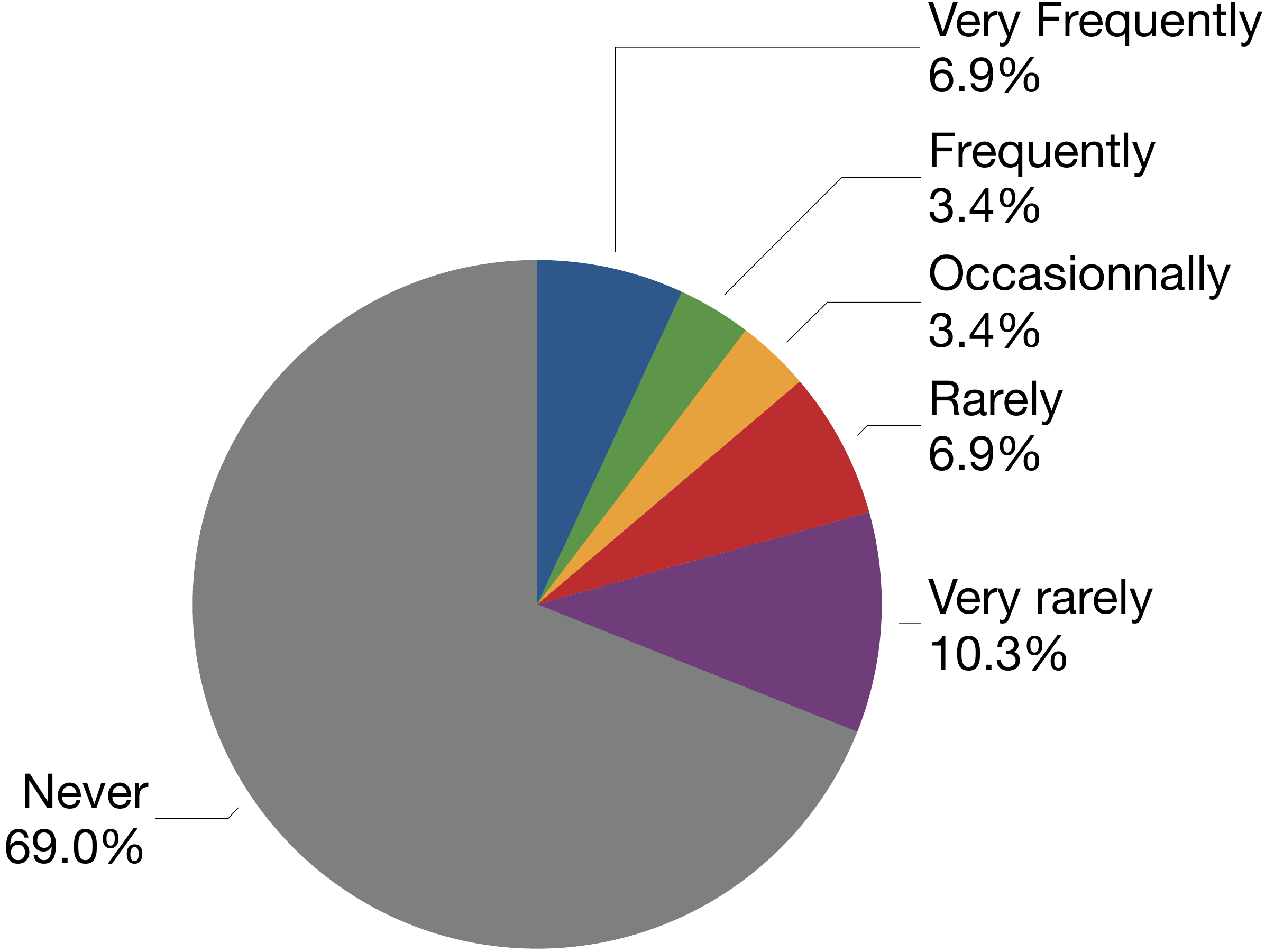}
		\caption{Group 1}
		\label{fig:survey_license_check_1}
	\end{subfigure}	\begin{subfigure}{.5\textwidth}
		\centering
		\includegraphics[width=.75\linewidth]{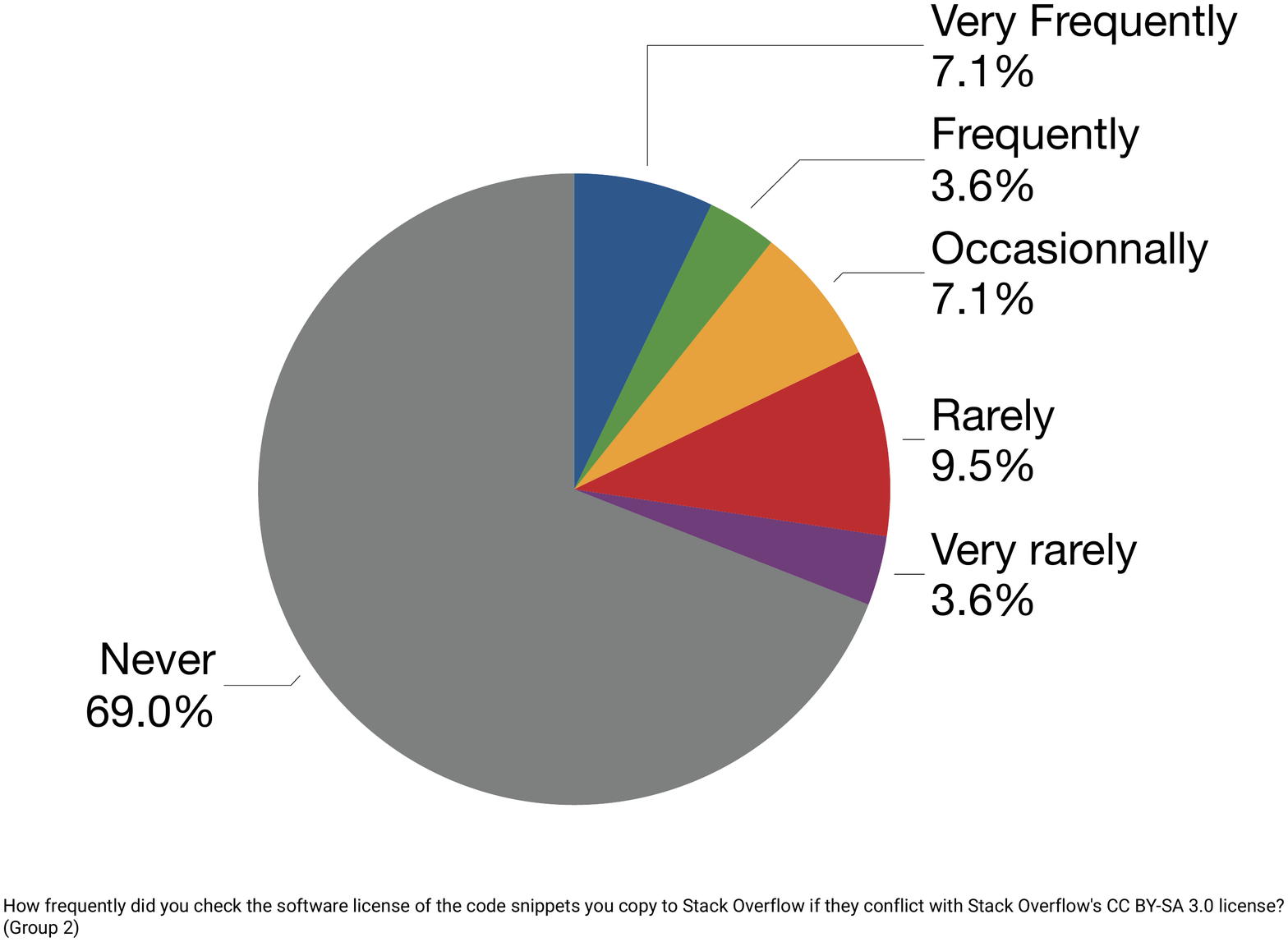}
		\caption{Group 2}
		\label{fig:survey_license_check_2}
	\end{subfigure}
	\caption{Frequency of the answerers checking license of their code snippets against Stack Overflow's CC BY-SA 3.0.}
	\label{fig:survey_license_check}
\end{figure}

We asked the answerers a follow-up question of how frequently they checked for
conflicts between software license of the code snippets they copied to their
answers and Stack Overflow's CC BY-SA 3.0. As shown in
Figure~\ref{fig:survey_license_check}, 80 (69\%) and 58 (69\%) answerers from
group 1 and group 2 did not perform the checking. There are only approximately
10\% of the answerers who frequently checks for licensing conflicts when they
copy code snippets to Stack Overflow.

\vspace{0.5cm} \noindent\fbox{	\parbox[c][1.5cm]{0.98\textwidth}{		\textit{For RQ3, approximately 62\% of our participants are aware of CC BY-SA
			3.0 license enforced by Stack Overflow. However, 98--99\% of the answerers
			never include software license in their Stack Overflow snippets. Sixty-nine
			percent never check for licensing conflicts when they copy code snippets to
			Stack Overflow answers.} }} \vspace{0.5cm}

To acquire additional insights, we invited every answerers for
further comments regarding their concerns of answering Stack Overflow (SO)
with code snippets. Some interesting comments are selected and discussed below.
The full set of answers can be found in the Appendix~\ref{appendixC}.

\begin{itemize}[label={}] \itemsep1em \item Comment 1: The answerer addresses a
	concern of programmers reusing his/her code snippets without understanding them.
	Moreover, he or she discusses a ramification of low-quality snippets or outdated
	code containing security issues on Stack Overflow.
	
	\begin{quote}\textit{The real issue is less about the amount the code snippets
			on SO than it is about the staggeringly high number of software
			``professionals'' that mindlessly use them without understanding what they're
			copying, and the only slightly less high number of would-be professionals that
			post snippets with built-in security issues.  A related topic is beginners who
			post (at times dangerously) misleading tutorials online on topics they actually
			know very little about. Think PHP/MySQL tutorials written 10+ years after
			\texttt{mysql\_*} functions were obsolete, or the recent regex tutorial that
			got posted the other day on HackerNew
			(\url{https://news.ycombinator.com/item?id=14846506}). They're also full of
			toxic code snippets.}\end{quote}
	
	\item Comment 2: The answerer suggests that a guidance from Stack Overflow
	regarding software license of code snippets will be beneficial.
	
	\begin{quote}\textit{When I copy code it's usually short enough to be
			considered ``fair use'' but I am not a lawyer or copyright expert so some
			guidance from SO would be helpful. I'd also like the ability to flag/review
			questions that violate these guidelines.}\end{quote}
	
	\item Comment 3: Similar to comment 1, the answerer addresses a concern of
reusing Stack Overflow code snippets without understanding.
	
	\begin{quote}\textit{My only concern, albeit minor, is that I know people
			blindly copy my code without even understanding what the code does.}\end{quote}
	
	\item Comment 4: As shown in RQ2, the answerer discusses a problem from outdated Stack Overflow code snippets and
	his/her solution.
	
	\begin{quote}\textit{The main problem for me/us is outdated code, esp. as old
			answers have high Google rank so that is what people see first, then try and
			fail. Thats why we're moving more and more of those examples to knowledge base
			and docs and rather link to those.}\end{quote}
	
	\item Comment 5: The answerers gives insights into the quality of the Stack
	Overflow code snippets.
	
	\begin{quote}\textit{Lot of the answers are from hobbyist so the quality is
			poor. Usually they are hacks or workarounds (even MY best answer on SO is a
			workaround).}\end{quote} \end{itemize}

\subsection{The visitor survey}

To answer RQ4 and RQ5, we used another online survey, the visitor survey, to ask
Stack Overflow visitors about their experiences of obsolete code and their
awareness to software license of Stack Overflow code snippets. We received 89
answers from 5 groups of Stack Overflow visitors. We combined the results and
present them as a single group below.

\subsubsection*{General Information}

\begin{figure}
		\centering
		\includegraphics[width=0.4\linewidth]{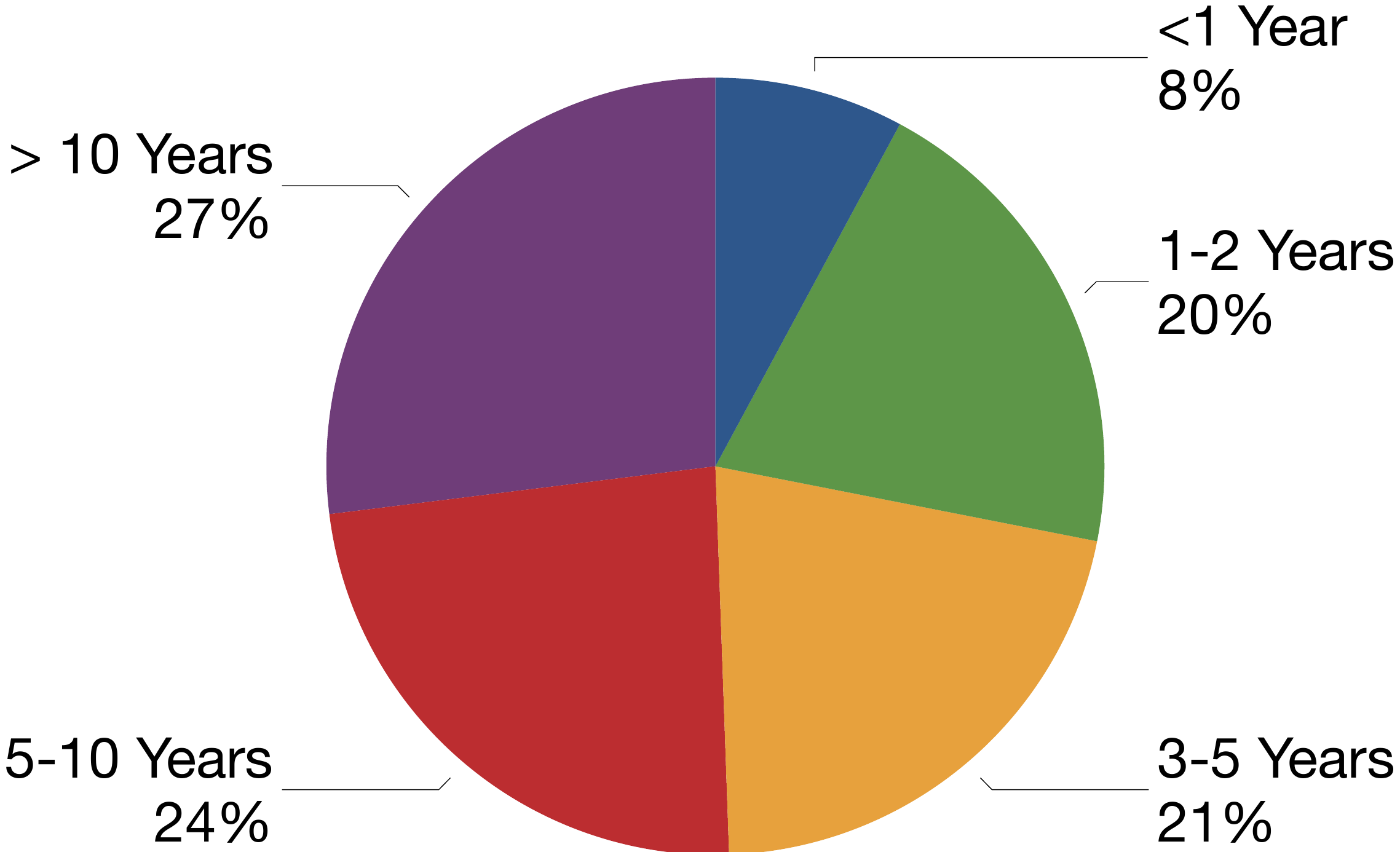}
		\caption{Experience of the Stack Overflow visitors}
		\label{fig:survey_visitor_exp}
\end{figure}

As illustrated in Figure~\ref{fig:survey_visitor_exp}, 24 (27\%) and 21 (24\%)
of the participants in Stack Overflow visitor survey have over 10 years and 5-10
years of experience respectively. There are 19 participants (21\%) who have 3-5
years, 18 (20\%) who have 1-2 years, and 7 (8\%) participants who have less than a
year of programming experience.

\subsubsection*{Where do the developers search for programming solutions?}

\begin{figure} \centering
	\includegraphics[width=.5\linewidth]{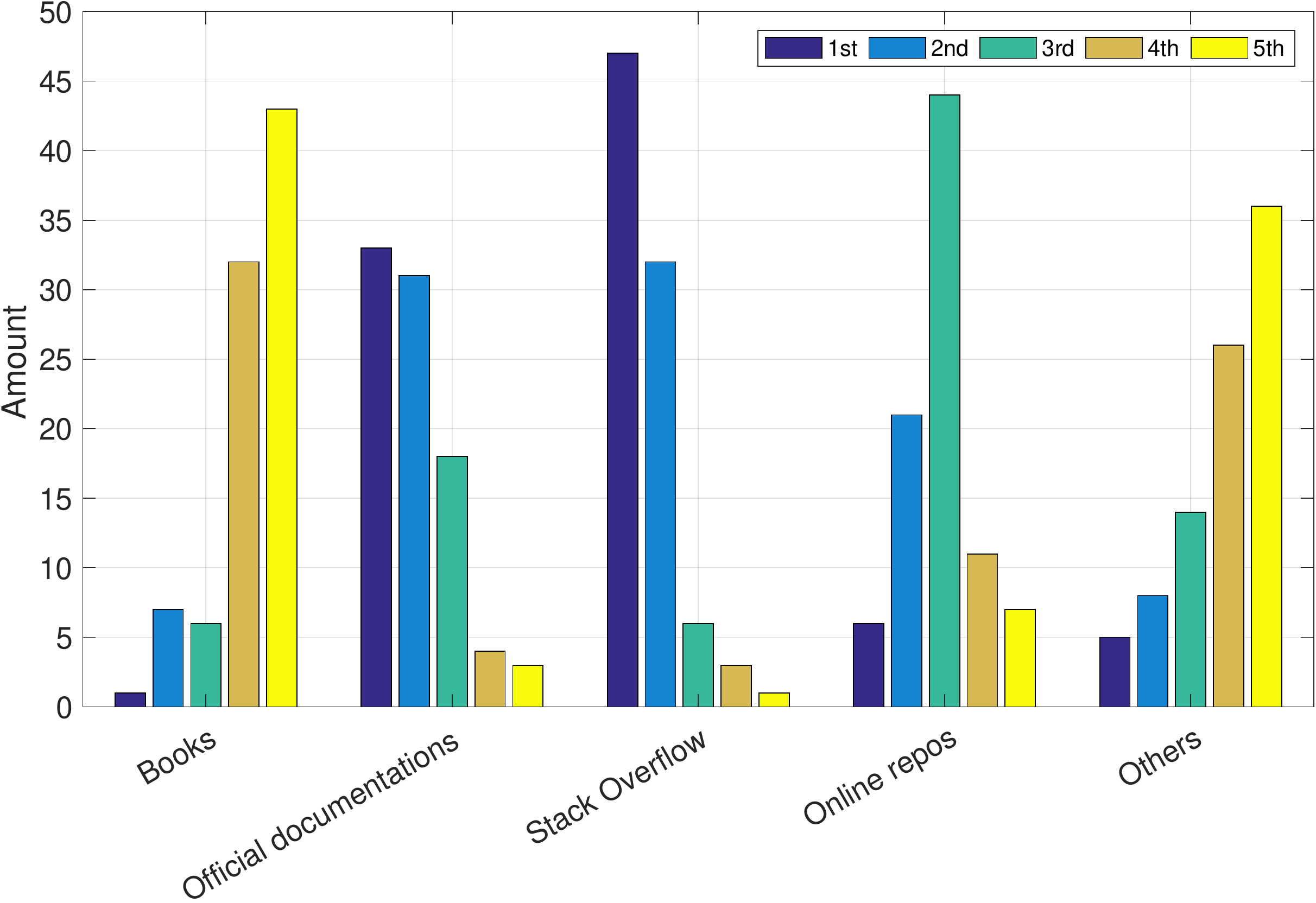} 
	\caption{Rankings of the resources developers use to solve programming problems}
	\label{fig:survey_visitor_rankings-crop} 
\end{figure}

We asked the participants to rank five options, without a tie, they will choose
to find programming solutions. The given five options include books, official
documentations, Stack Overflow, online repositories (e.g.\ GitHub), and others.
The results are displayed in Figure~\ref{fig:survey_visitor_rankings-crop}. We
found that 47 out of 89 participants rank Stack Overflow as the 1st option to
search for programming solutions, followed by official documentation (33),
online repositories (6), other resources (5), and books (1).

\begin{figure} \centering
	\includegraphics[width=.5\linewidth]{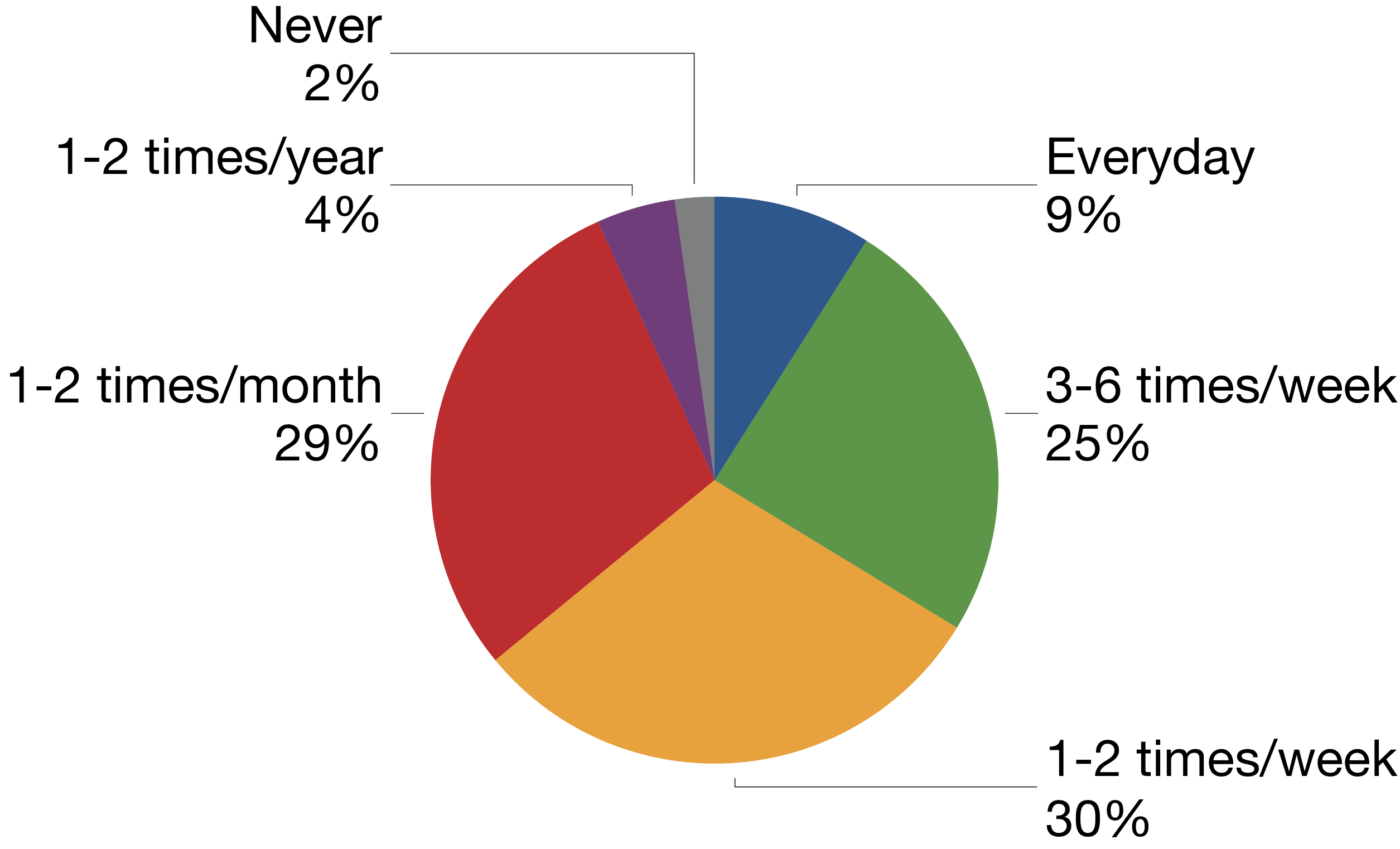} 
	\caption{Frequency of copying code from Stack Overflow}
	\label{fig:survey_visitor_frequency_so_copy-crop} 
\end{figure}

Since Stack Overflow is among the first research to solve programming tasks, we
asked the participants how frequently they reuse code snippets from answers on
Stack Overflow. According to the results (see
Figure~\ref{fig:survey_visitor_frequency_so_copy-crop}), we found that 57 (64\%)
participants actively reusing code snippets from Stack Overflow. Eight
participants (9\%) copy Stack Overflow code every day, 22 (25\%) do copying 3-6
times a week, 27 (30\%) do copying once or twice a week. There are 2
participants (2\%) who never copy code from Stack Overflow.

\begin{figure} \centering
	\includegraphics[width=0.6\linewidth]{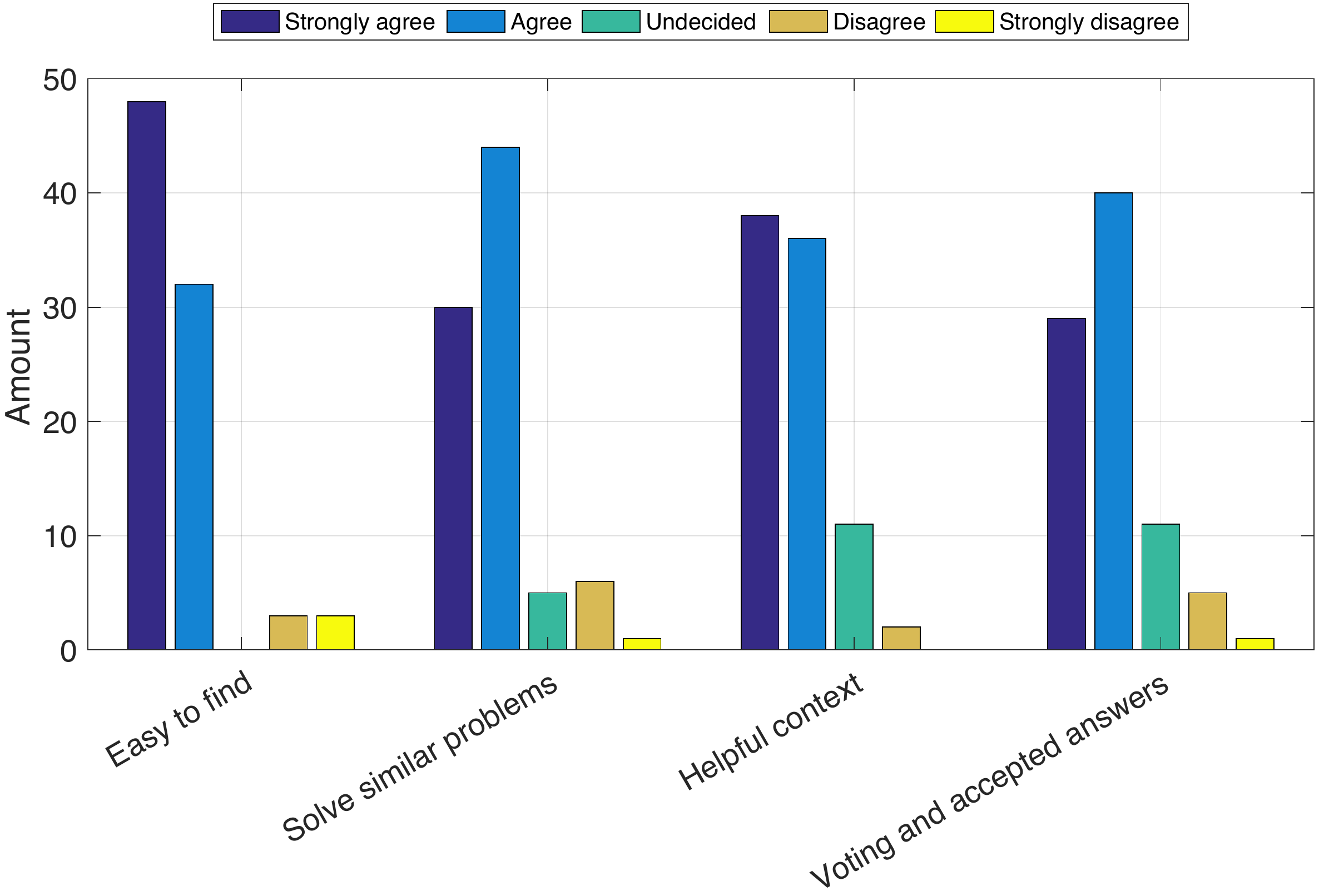} 
	\caption{Why did you copy and reuse code snippets from Stack Overflow?}
	\label{fig:survey_visitor_why_copy_so} 
\end{figure}

To understand why the participants choose to copy code snippets from Stack
Overflow, we asked them to rate four reasons in Likert's scale (Strongly agree,
Agree, Undecided, Disagree, Strongly disagree). The four reasons include
\textit{They are easy to find by searching the web}, \textit{They solve problems
	similar to my problems with minimal changes}, \textit{The context of questions
	and answers helped me understand the code snippets better}, \textit{The voting
	mechanism and accepted answers helped to filter good code from bad code}. The
answers are depicted in Figure~\ref{fig:survey_visitor_why_copy_so}. More than
80\% of the participants agree with all the four reasons. We observed only two
``disagree'' and zero ``Strongly disagree'' answer for ``Helpful context'', the
lowest disagreement among the four reasons. This means most of them agree that
the context of questions and answers on Stack Overflow help them understand the
code snippet better.

To sum up, Stack Overflow is ranked higher than official documentation, online
repositories, and books as the resource to look for programming solutions.
Developers rely on Stack Overflow answers because they are easy to search for on
the web. Moreover, 64\% of the participants reuse code snippets from Stack
Overflow at least once a week. The copy code from Stack Overflow because they
can be found easily from search engine, solve similar problems to their
problems, provide helpful context, and offer voting and accepted answers.

\subsubsection*{RQ4: What are the problems Stack Overflow visitors experiencing
	from reusing code snippets on Stack Overflow?}

\begin{figure} \centering
	\includegraphics[width=.4\linewidth]{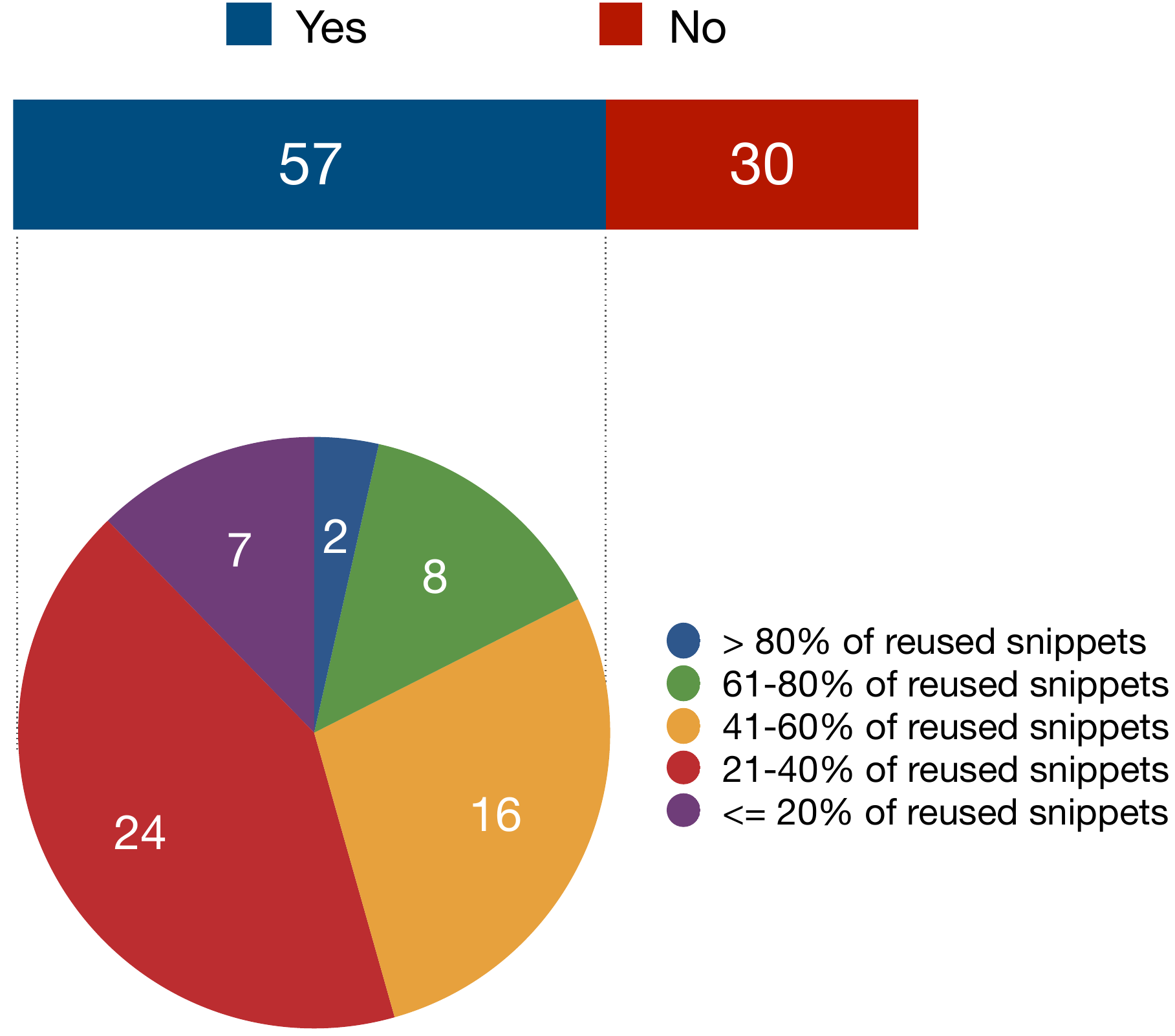} 
	\caption{Number of visitors who experienced a problem from reusing Stack Overflow code snippets and the frequency.}
	\label{fig:survey_v_problems} 
\end{figure}

We asked the visitors whether they have had any problem from reusing Stack Overflow
code snippets and how often did the problems occur.  Fifty seven out of eighty
seven participants (66\%) experienced a
problem from reusing Stack Overflow snippets (see Figure~\ref{fig:survey_v_problems}). Among the 57, there are 2
participants who found problems in more than 80\% of the reused code snippets.
Eight and sixteen faced problems from at least sixty and forty percent of the
reused snippets.

\begin{figure} \centering
	\includegraphics[width=.5\linewidth]{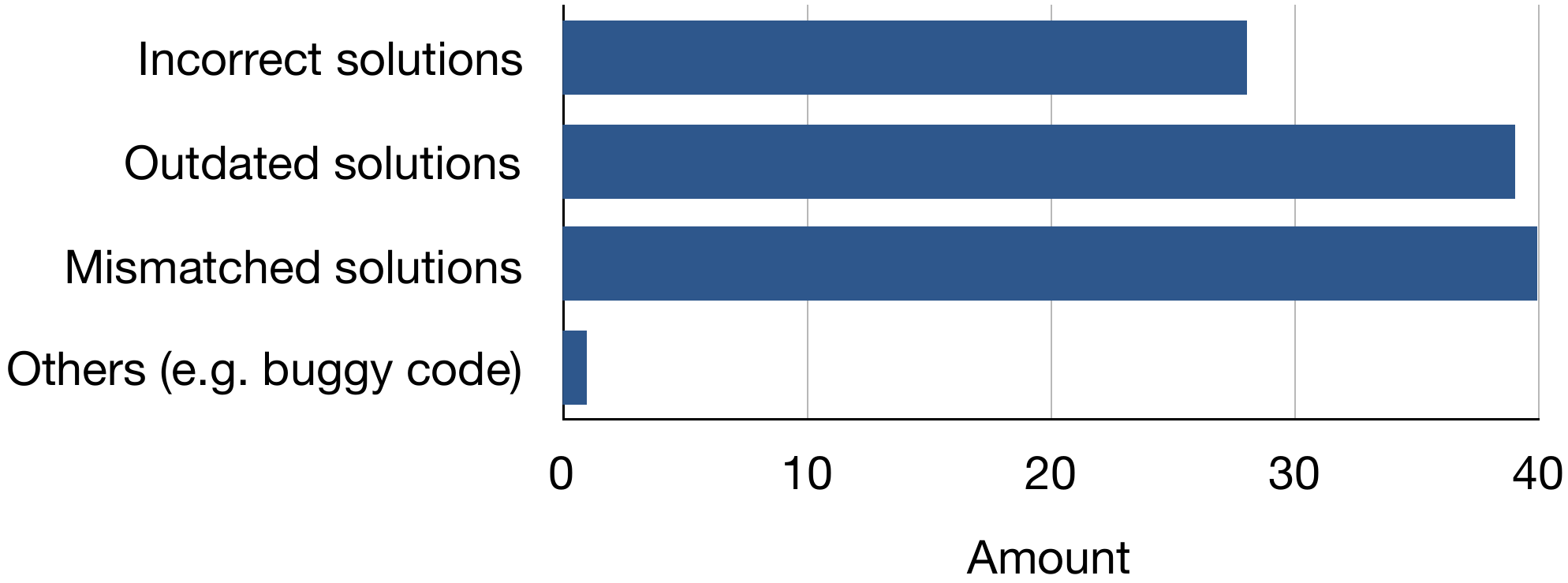} 
	\caption{Problems from reusing Stack Overflow code snippets}
	\label{fig:survey_visitor_problems} 
\end{figure}

The problems from reusing Stack Overflow code snippets (illustrated in
Figure~\ref{fig:survey_visitor_problems}) include incorrect solutions, i.e.~the
code claims to solve the problem in the question while it does not  (28 out of
57 $\approx$ 49\% participants reported this); outdated solutions, i.e.~the code
may work with the some older versions of the library or API, but not the one
they are using (39 out of 57 $\approx$ 68\% participants reported this);
mismatched solutions, i.e.~the code solves the problem in the question but it is
not exactly the right solution for their problem (40 out of 57 $\approx$ 70\%
participants reported this); and buggy code (1 out of 57 $\approx$ 2\%
participants reported this).

\begin{figure} \centering
	\includegraphics[width=.4\linewidth]{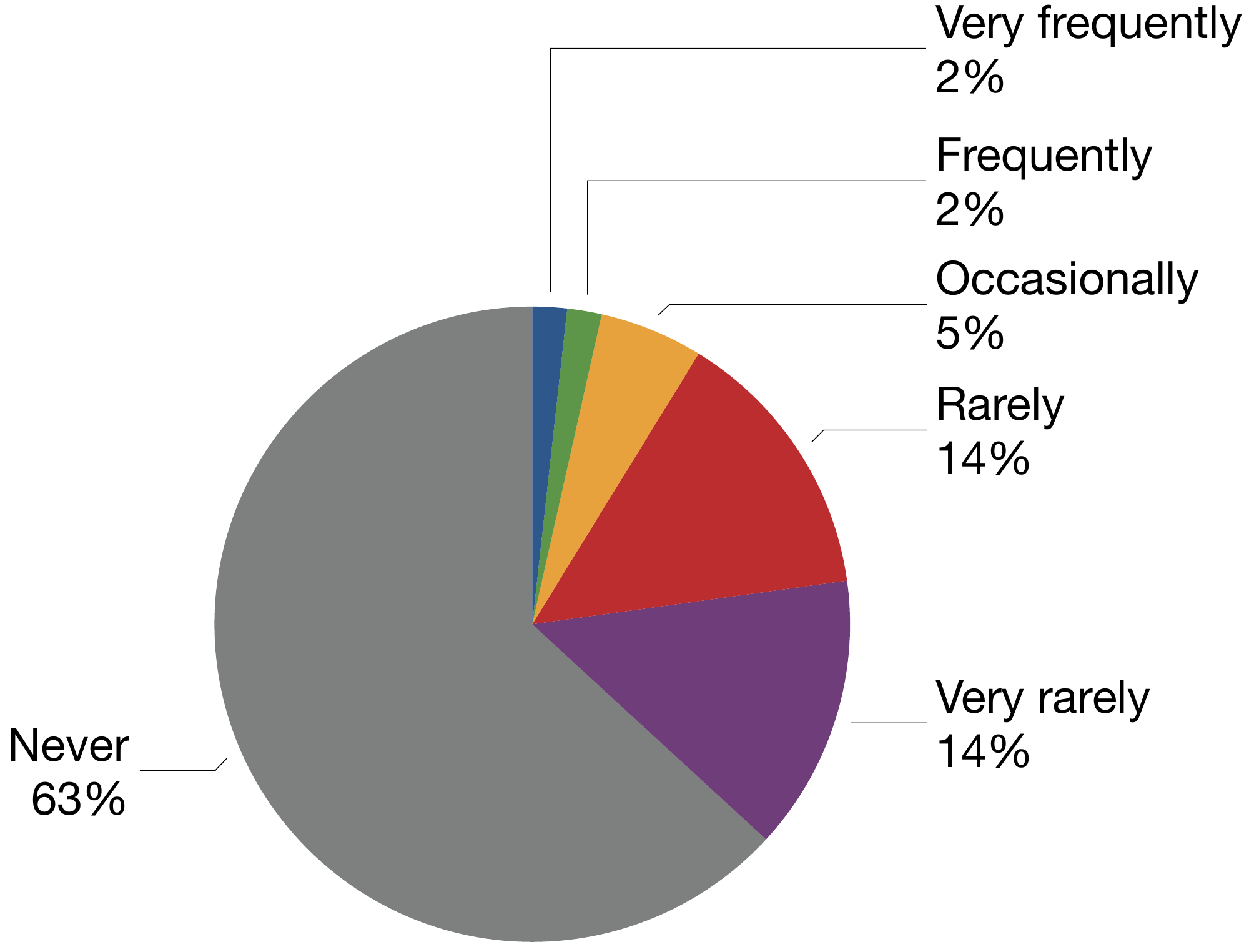} 
	\caption{Frequency of Stack Overflow visitors reporting the problems back to Stack Overflow discussion threads}
	\label{fig:survey_visitor_freq_report_problems} 
\end{figure}

\begin{figure} \centering
	\includegraphics[width=.5\linewidth]{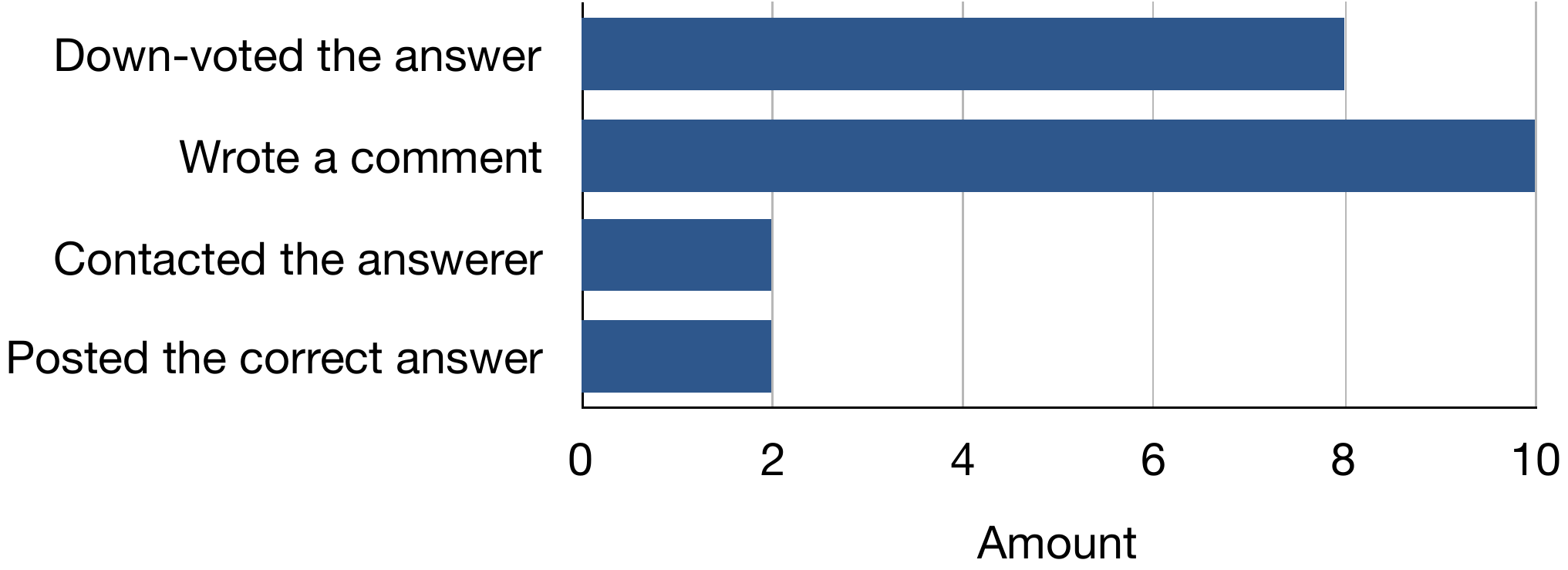} 
	\caption{Options that Stack Overflow visitors choose to report the problems from Stack Overflow snippets}
	\label{fig:survey_visitor_how_report_problems} 
\end{figure}

Stack Overflow visitors rarely report the problems back to the discussion
threads (as can be seen in
Figure~\ref{fig:survey_visitor_freq_report_problems}). Among the 57 participants
who encounter problems of Stack Overflow snippets, 36 of them (63.2\%) never
report the problems. Fourteen participants who reported the problems
did so by writing a comment (10), down-voting the answer (8), contacted the
answerer (2), and posting the new and correct answer (2) (shown in
Figure~\ref{fig:survey_visitor_how_report_problems}).

\vspace{0.5cm} \noindent\fbox{	\parbox[c][2.2cm]{0.98\textwidth}{		\textit{For RQ4, our survey results show that 57 out of 87 Stack Overflow
			visitors encountered a problem from reusing Stack Overflow code snippets. Ten
			participants experienced problems for more than 80\% of the copied snippets and
			sixteen participants faced problems for 40--60\% of the reused code. The
			problems ranked by frequency include mismatched solutions (40), outdated
			solutions (39), incorrect solutions (28), and buggy code (1). Sixty-three percent
			of the participants never report the problems back to Stack Overflow.}}} \vspace{0.5cm}

\subsubsection*{RQ5: Are Stack Overflow visitors aware of code licensing on Stack Overflow?}

\begin{figure} \centering
	\includegraphics[width=.4\linewidth]{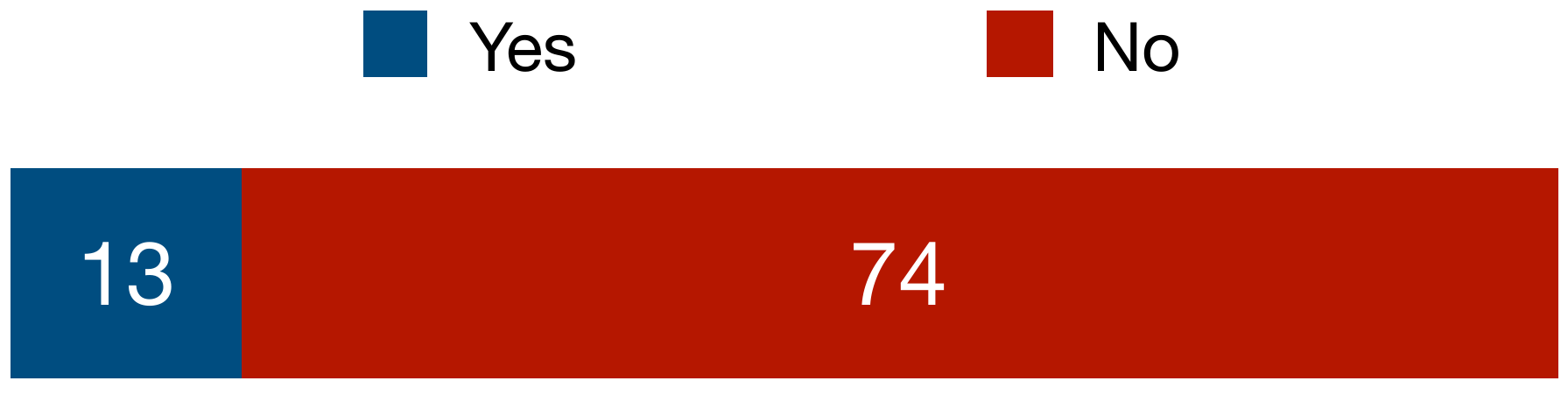} 
	\caption{Awareness of Stack Overflow visitor to CC BY-SA 3.0 license}
	\label{fig:survey_visitor_cc_by-sa} 
\end{figure}

\begin{figure} \centering
	\includegraphics[width=.4\linewidth]{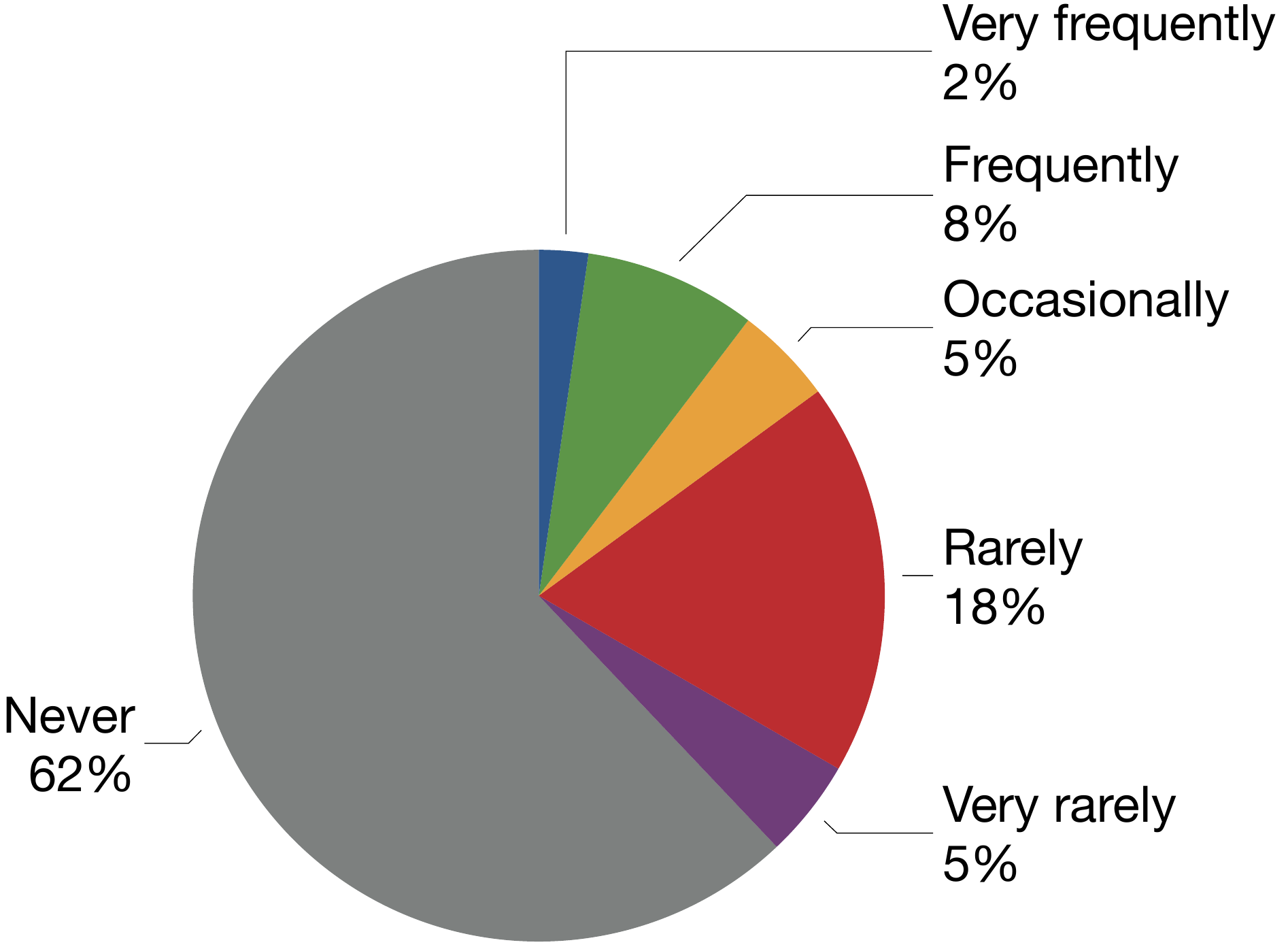} 
	\caption{Attributions to Stack Overflow when reusing code snippets}
	\label{fig:survey_visitor_attribution-sa} 
\end{figure}

As depicted in Figure~\ref{fig:survey_visitor_cc_by-sa}, 74 out of 87 (85\%)
Stack Overflow visitors are not aware, at the time of copying the code, that
Stack Overflow apply Creative Commons Attribution-ShareAlike 3.0 Unported (CC
BY-SA 3.0) to content in the posts, including code snippets. As a consequence
62\% of the visitors never give an attribution, which is required by CC BY-SA
3.0, to a Stack Overflow post they copied the code from (the complete statistics
can be found from Figure~\ref{fig:survey_visitor_attribution-sa}).

\begin{figure} \centering
	\includegraphics[width=.4\linewidth]{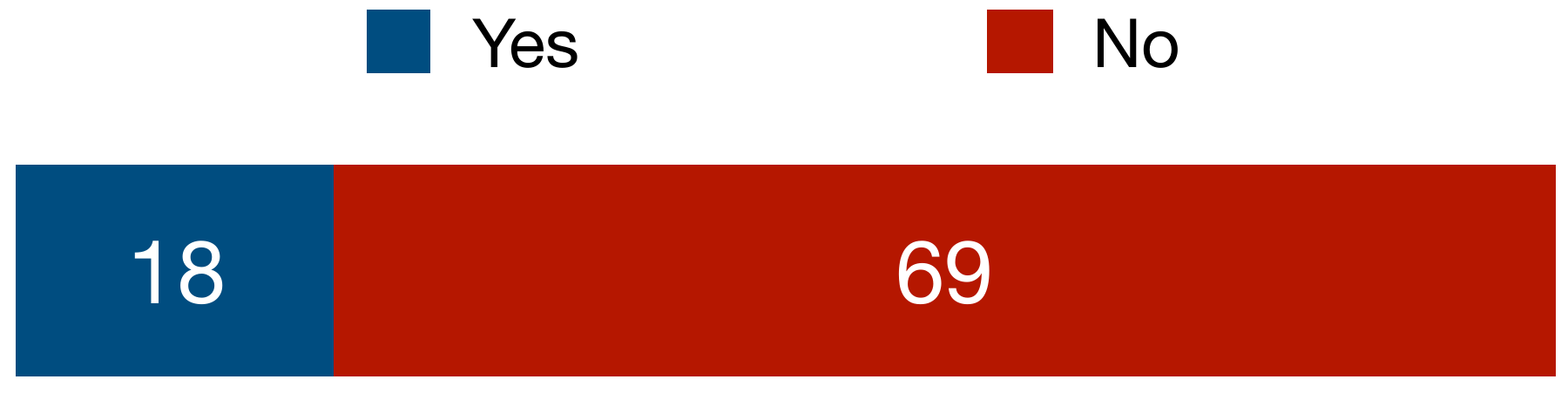} 
	\caption{Checking for the original license of Stack Overflow code snippets}
	\label{fig:survey_visitor_original_license} 
\end{figure}

Sixty-nine Stack Overflow visitors (79\%) who adopted code from Stack Overflow
never check if the code snippet originated from a different source (e.g.\ an open
source project) with an incompatible license to their projects (see
Figure~\ref{fig:survey_visitor_original_license}).

\begin{figure} \centering
	\includegraphics[width=.4\linewidth]{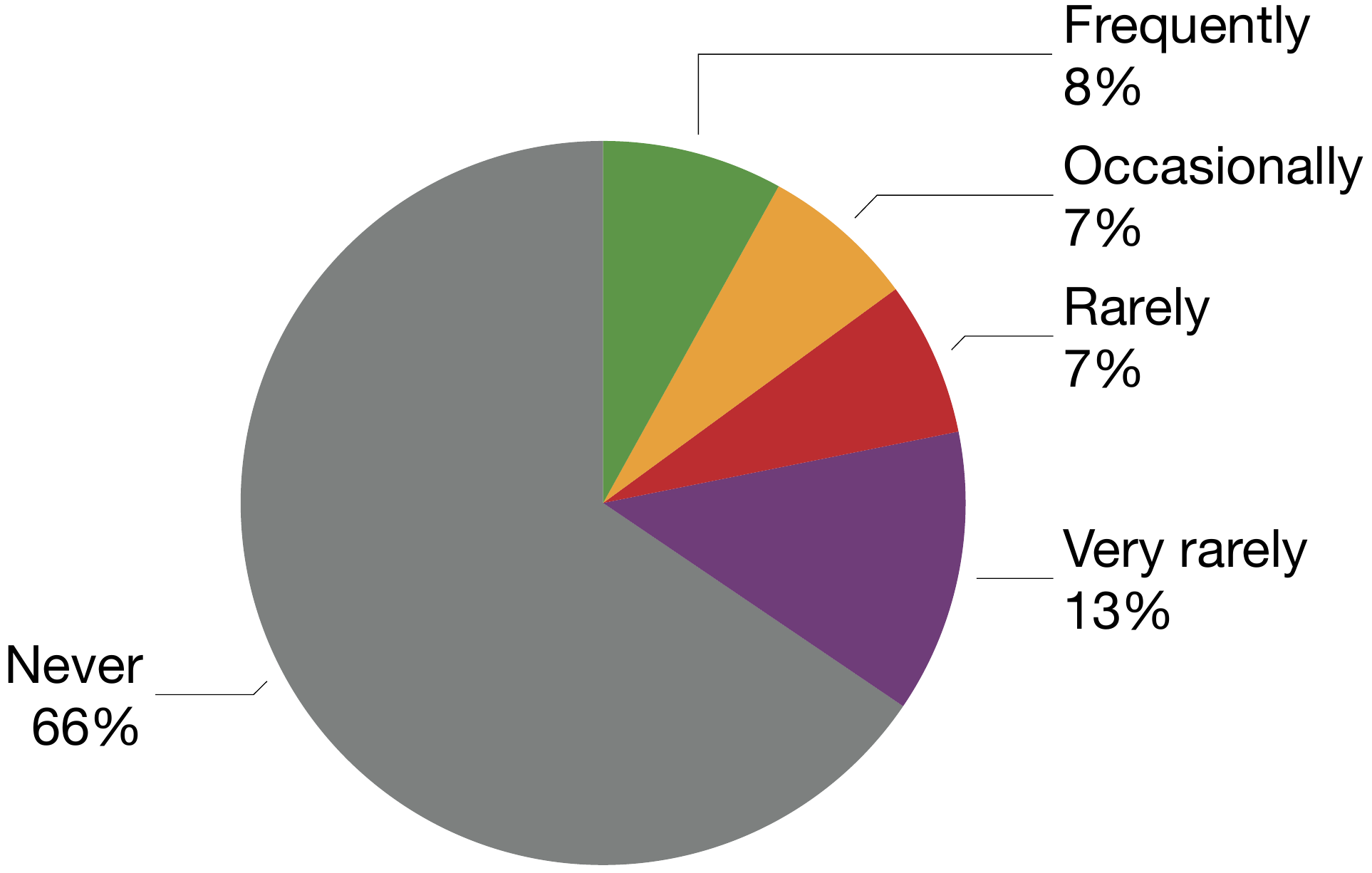} 
	\caption{Checking for licensing conflicts from reusing Stack Overflow snippets}
	\label{fig:survey_visitor_licensing_conflict_check} 
\end{figure}

\begin{figure} \centering
	\includegraphics[width=.4\linewidth]{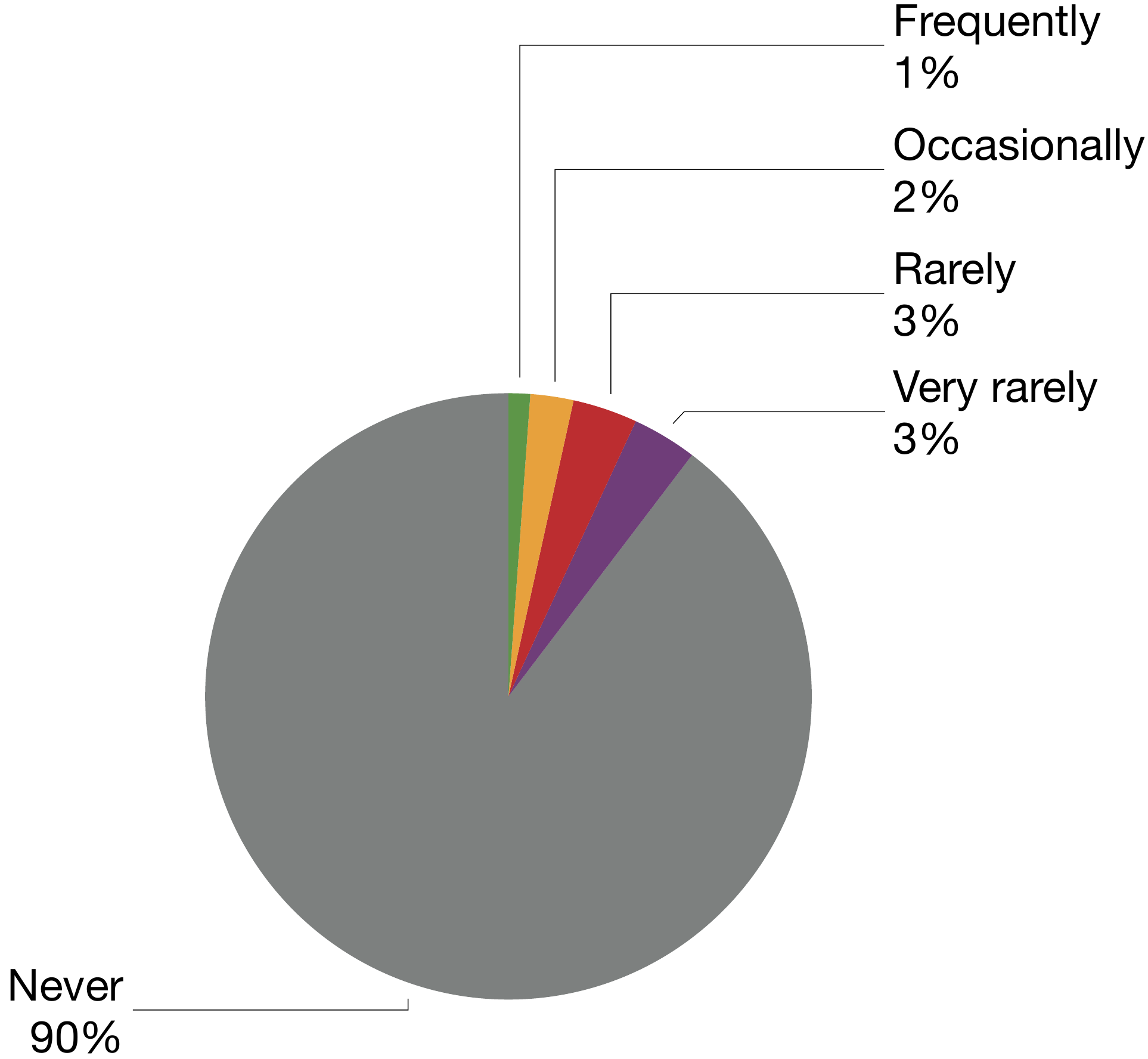} 
	\caption{Legal issues found from reusing Stack Overflow code snippets}
	\label{fig:survey_visitor_legal_issue} 
\end{figure}

Fifty seven participants (66\%) never check for licensing conflicts at all
when reusing Stack Overflow code (see
Figure~\ref{fig:survey_visitor_licensing_conflict_check}). Lastly, 9\% of
the participants experienced legal issues by reusing code snippets on Stack
Overflow (see Figure~\ref{fig:survey_visitor_legal_issue}). We did not
expect that any participant encountered legal issues as we are not
aware of such cases being reported in the literature. It would
be interesting to followup on the kind of legal issues that have been
encountered, however, as we designed the
survey to be anonymous, it was not possible to contact the participants for further details.

\vspace{0.5cm} \noindent\fbox{	\parbox[c][1.7cm]{0.98\textwidth}{		\textit{For RQ5, 85\% of the participants are not aware of Stack Overflow CC
			BY-SA 3.0 license and 62\% never give attributions to the Stack Overflow posts
			they copied the code snippets from. We found that 66\% of the visitors never check for software
	licensing conflicts between Stack Overflow code snippets and their projects.
	9\% of the participants encountered legal issues.}}} \vspace{0.5cm}

\subsection{Overall Discussion}

By separating the answerers into two groups according to their reputation, we
observed some similarities and differences in their responses. The sources of
code snippets in Stack Overflow answers are similar in both groups. The
answerers mainly write the code snippets from scratch aiming to answer the
question. The frequencies of copying from each source, i.e.~personal projects,
company projects, open source projects, writing from scratch, modifying from the
questions, and others, also follow the same proportions for both group.

The main difference we found is responses regarding outdated code. The answerers
in Group 1 have a more substantial percentage (61.2\%) of being notified about
the outdated code in their answers than Group 2 (47.1\%). This may be caused by
the amount of their Stack Overflow answers. Since Group-1 answerers have a
higher reputation than Group-2 answerers, they possibly have given more answers
on Stack Overflow. The higher amount of answers hence increase the chance of the
code being outdated. Interestingly, although the percentage of outdated code
notifications in Group 2 is lower, the percentage of the answerers who very
frequently fixed their outdated code is higher than of Group 1.

Regarding software license, the responses from the two groups agree with each
other. The answerers in both groups show the same level of awareness to Stack
Overflow CC BY-SA 3.0 license (62.1\% and 62.3\%). Similarly, the answerers in
both groups neither include software license in their code snippets (98\% and
99\%) nor check of licensing conflicts between their code snippet's and Stack
Overflow CC BY-SA 3.0 (69\%).

The visitors survey confirms the findings from the previous studies that Stack
Overflow code snippets can be problematic~\citep{Acar2016,An2017}. Sixty-six
percent of the visitors experienced a problem from reusing Stack Overflow code
snippets ranging from incorrect solutions, outdated solutions, mismatched
solutions, to buggy code. Although they are aware of the problems, half of them
(56\%) never reports back to the Stack Overflow discussions. On the other hand,
the visitors rarely give attributions to Stack Overflow when they reuse code
snippets from the website, similar to the findings reported
by~\cite{Baltes2017}. The visitors are generally not aware of the CC BY-SA 3.0
license, and more than half of them never check for license compatibility when
reusing Stack Overflow code snippets. We also found that 9\% of the participants
encountered legal issues by copying code from Stack Overflow. To the best of our
knowledge, a study of legal problems from reusing Stack Overflow code snippets
has never been done before and will be our future work.

\section{Threats to Validity}

\subsection{Internal Validity}
We only invited 607
developers to participate in the answerer survey. If all the developers are
invited, which is almost impossible considering 7.6 million users on Stack
Overflow, the results may be different. Nevertheless, we mitigate this threat to
internal validity by selecting the participants based on their Stack Overflow
reputation. This targeted participant selection ensures that we
invite developers who have been actively involved in asking and answering
questions on the site for a long period of time, and refrain from inviting new
users who merely answer a question or two. The highest reputation user invited
to our survey answered 33,933 questions and the lowest reputation user in our
survey answered 116 questions.

We selected the participants for the visitor survey based on convenient sampling
which could suffer from bias and outliers. We mitigate the threat by inviting
different groups of participants ranging from the author's social media,
technology news and media community, software engineering community (on
Facebook), Java programmer discussion thread (com.lang.java.programmer), and
from University of Molise.

\subsection{External Validity} 
While the reputation is a good proxy to reflect the amount of answers the
developers have given, it might not cover all kinds of answerers and their
experience on Stack Overflow. The answerers who have lower-reputation than 6,999
were not invited to our study and their awareness and experience may differ from
our findings. Hence, our findings may not be generalised to all Stack Overflow
answerers.

At least 39\% of the participants in the visitor survey are from Thailand
(\textsf{blognone.com} and some of the first author's contacts). Due to working
culture, answers from this group of developers may only represent software
developers in Thailand. Similarly our findings from the visitor survey may not
be generalised to all Stack Overflow visitors. We alleviate this concern by
inviting participants from other groups, e.g.~University of Molise in Italy,
comp.lang.java.programmer, and Software Engineering Facebook group.

\section{Related Work}

\subsection{Stack Overflow}

Stack Overflow is a gold mine for software engineering research and has been put
to use in several previous studies. In terms of developer-assisting tools,
Seahawk is an Eclipse plug-in that searches and recommends relevant code
snippets from Stack Overflow~\citep{Ponzanelli2013}. A follow-up work, Prompter,
by Ponzanelli et al.~\citep{Ponzanelli2014} achieves the same goal but with
improved algorithms. The code snippets on Stack Overflow are mostly examples or
solutions to programming problems. Hence, several code search systems use whole
or partial data from Stack Overflow as their code search
databases~\citep{Keivanloo2014,Park2014,
	Stolee2014,Subramanian2013,Diamantopoulos2015}. Furthermore, Treude et
al.~\cite{Treude2016}~use machine learning techniques to extract insight
sentences from Stack Overflow and use them to improve API documentation.

Another research area is knowledge extraction from Stack Overflow.
\cite{Nasehi2012}~studied what makes a good code example by analysing answers
from Stack Overflow. Similarly, \cite{Yang2016} report the number of reusable
code snippets on Stack Overflow across various programming languages.
\cite{Wang2013_StackOverflow} use Latent Dirichlet Allocation (LDA) topic
modelling to analyse questions and answers from Stack Overflow so that they can
automatically categorise new questions. There are also studies trying to
understand developers' behaviours on Stack Overflow, e.g.~A study by
\cite{Movshovitz-Attias2013,Rosen2016,Choetkiertikul2015,Bosu2013}.
\cite{Yang2017} analysed 909k non-fork Python projects on GitHub and 1.9 million
python code snippets on Stack Overflow and found thousands of code blocks that
are copied from Stack Overflow to GitHub. \cite{Baltes2017} discovered that two
thirds of the clones from the 10 most frequently referenced Java code snippets on Stack
Overflow do not contain attributions.

\subsection{Software Licensing} Software licensing is crucial for open source
and industrial software development. Di Penta et al.~\cite{DiPenta2010} studied
the evolution of software licensing in open source software and found that
licensing statements change over time. \cite{German2009} found that licensing
incompatibility occur between the clone siblings, i.e.~clones among different systems
that come from the same source. Later, \cite{German2010} created an automated
tool for software license identification, Ninka. \cite{An2017} detected clones
between 399 Android apps and Stack Overflow Java code snippets. They found  that
1,279 cloned snippets potentially violate software licenses. Our study gives
insights into the software licensing of Stack Overflow code snippets. We study
the licensing awareness of developers when they copy and paste code to/from
Stack Overflow.

\subsection{Reusing of outdated third-party source code} Outdated code occurs in
software development. Xia et al.~\cite{Xia2014} discover that there is code
reused from popular open source projects in a large number of open source
systems. More importantly, some of the reused code is outdated which affect
quality and security of the software. Our study focuses on a similar problem but
on the context of Stack Overflow.

\section{Conclusions}

We present the results from online surveys of Stack Overflow answerers and
visitors regarding their awareness and experience of answering and reusing code
snippets on Stack Overflow. We are particularly interested in cloned code
snippets in Stack Overflow answers. Thus, the two online surveys are performed to
find out the \textit{origin} of code snippets in Stack Overflow answers and
\textit{awareness} of Stack Overflow developers, both answerers and visitors,
to the problem of outdated and license-violating code.

From the answerer's responses, we discover that code in Stack Overflow answers
is usually written from scratch to answer a question. Nonetheless, there are
some code snippets that are copied from the answerer's personal or company
projects, or from open source software.

Outdated code on Stack Overflow occurs when a code snippet is
copied from another location and is no longer up-to-date. The results show that
although the answerers are aware of outdated code in their answers, 19\% of them
rarely or never fix the code. One answerer expresses a concern of having
outdated code in his/her popular answer reused
by many other developers.

Approximately 60\% of the answerers are aware that Stack Overflow
applies the CC BY-SA 3.0
license to code on Stack Overflow. Ninety-eight to ninety-nine percent
do not include a license in their answers nor check for licensing conflicts when
they copy the code from another location to their answer. Some developers
believe the code is too small and fall under fair use policy. Many answerers
mention the license of their code in their profile page instead.

From the visitor's responses, Stack Overflow is ranked the highest as a resource
to solve programming tasks. They agree that Stack Overflow answers are easy to
find by search engines. More than half of them (64\%) reuse code from Stack
Overflow at least once a week. Sixty-seven percent of the visitors find a
problem from reusing Stack Overflow snippets ranging from a mismatched solution,
outdated solution, incorrect solution, and buggy code. However, 56\% of the
participants never report the problems back to Stack Overflow.

Eighty-five percent of the visitors are not aware of Stack Overflow's CC BY-SA
3.0 and sixty-two percent never give attributions to Stack Overflow posts that they copied
the code from. Moreover, sixty-six percent never check for licensing conflicts while
reusing Stack Overflow code snippets in their software
projects. Unexpectedly, 9\%
of the participants encountered legal issues.

With these findings, we
suggest Stack Overflow raise an awareness of their users, both the answerers and
the visitors, to the problem of outdated and license-violating code snippets.

\bibliographystyle{spbasic}
\bibliography{sigproc} 

\clearpage
\section{Appendix}

This appendix contains additional materials of the Stack Overflow answerer
survey questions, the visitor survey questions, and the complete set of open
comments from the Stack Overflow answerers.

\clearpage

\appendix
\section{Stack Overflow Answerer Survey}\label{appendixA}
\begin{figure}[H]
	\centering
	\includegraphics[width=0.8\linewidth]{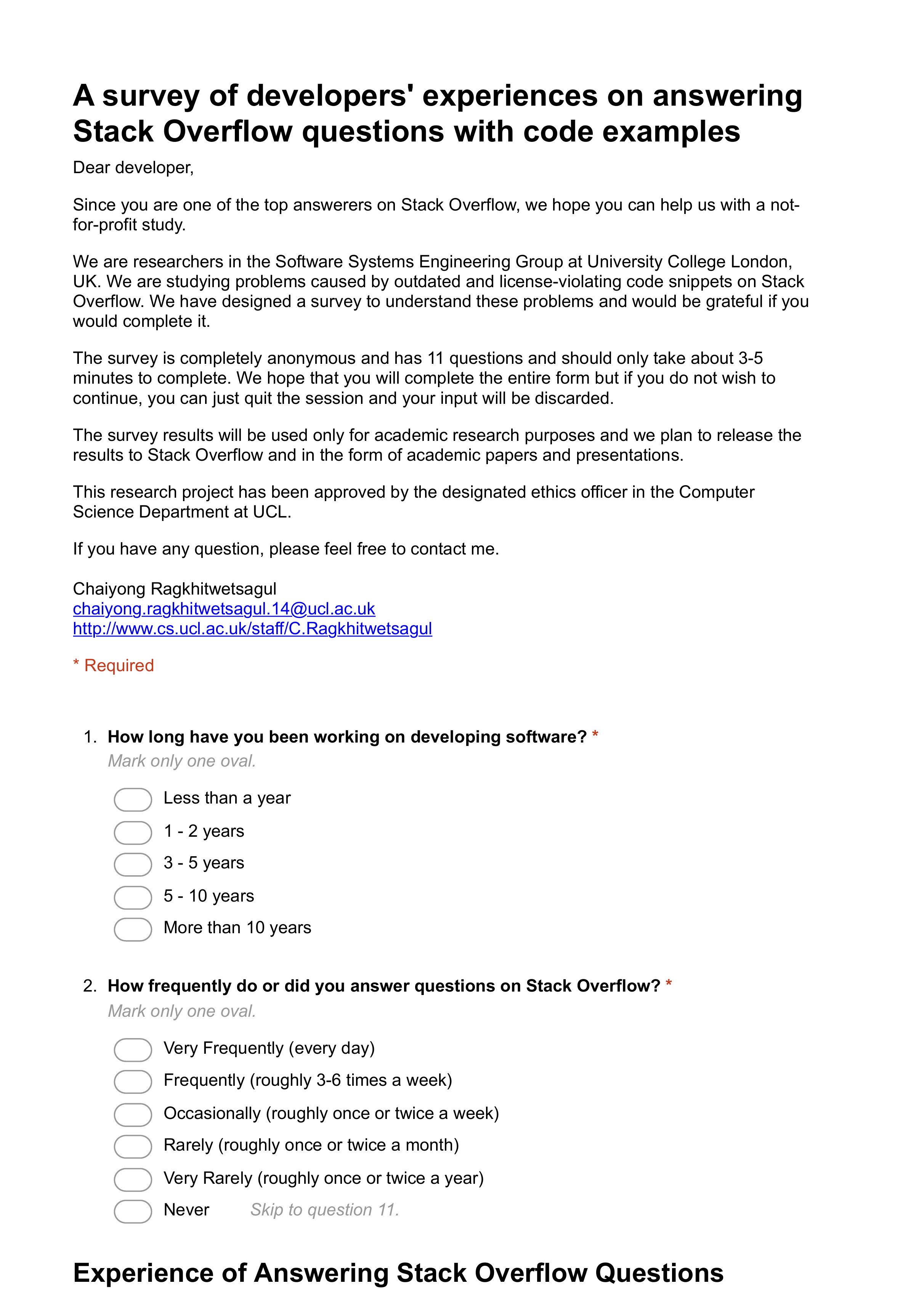}
	\label{fig:answerer-1}
\end{figure}

\begin{figure}[H]
	\centering
	\includegraphics[width=0.9\linewidth]{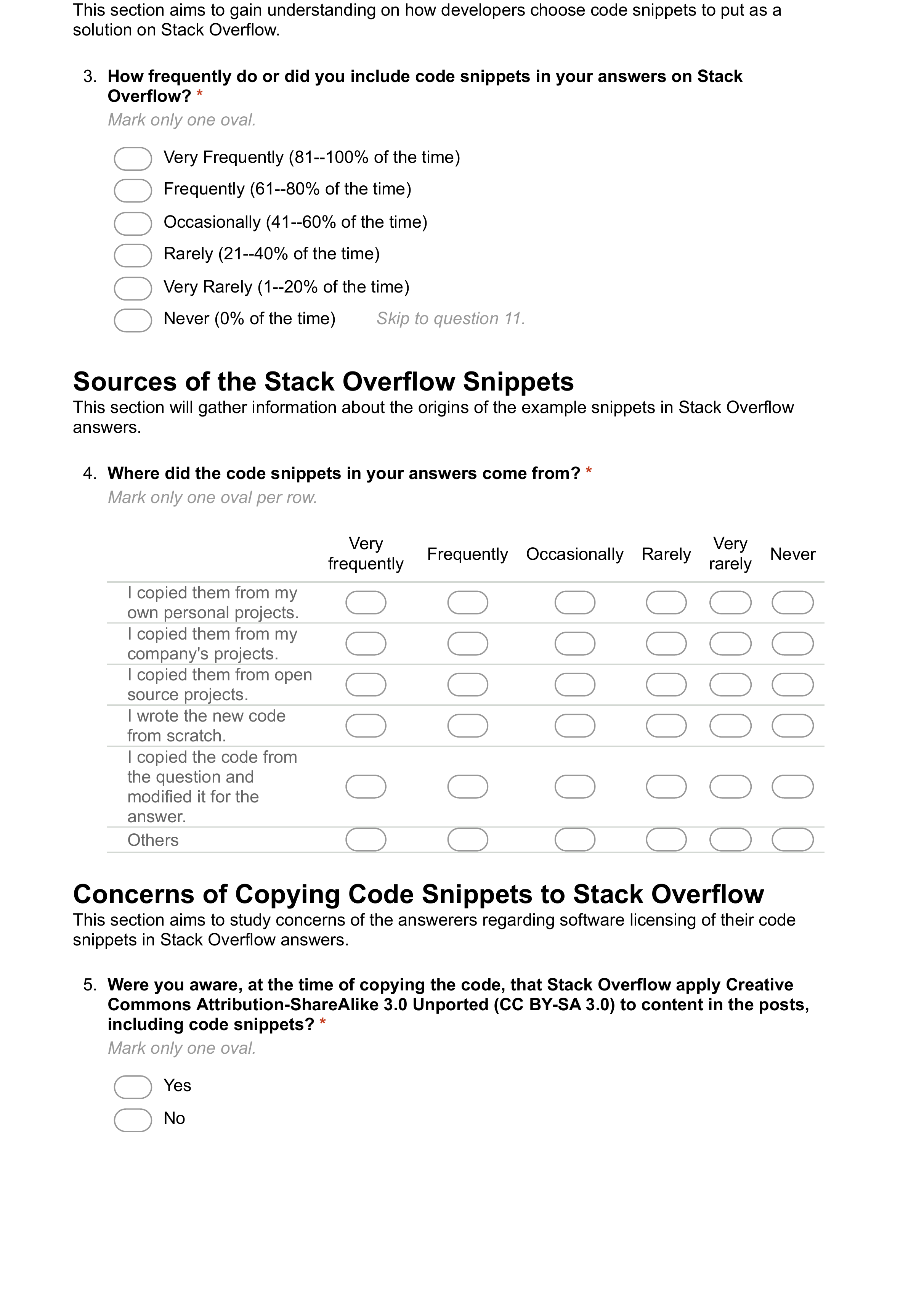}
	\label{fig:answerer-2}
\end{figure}

\begin{figure}[H]
	\centering
	\includegraphics[width=0.9\linewidth]{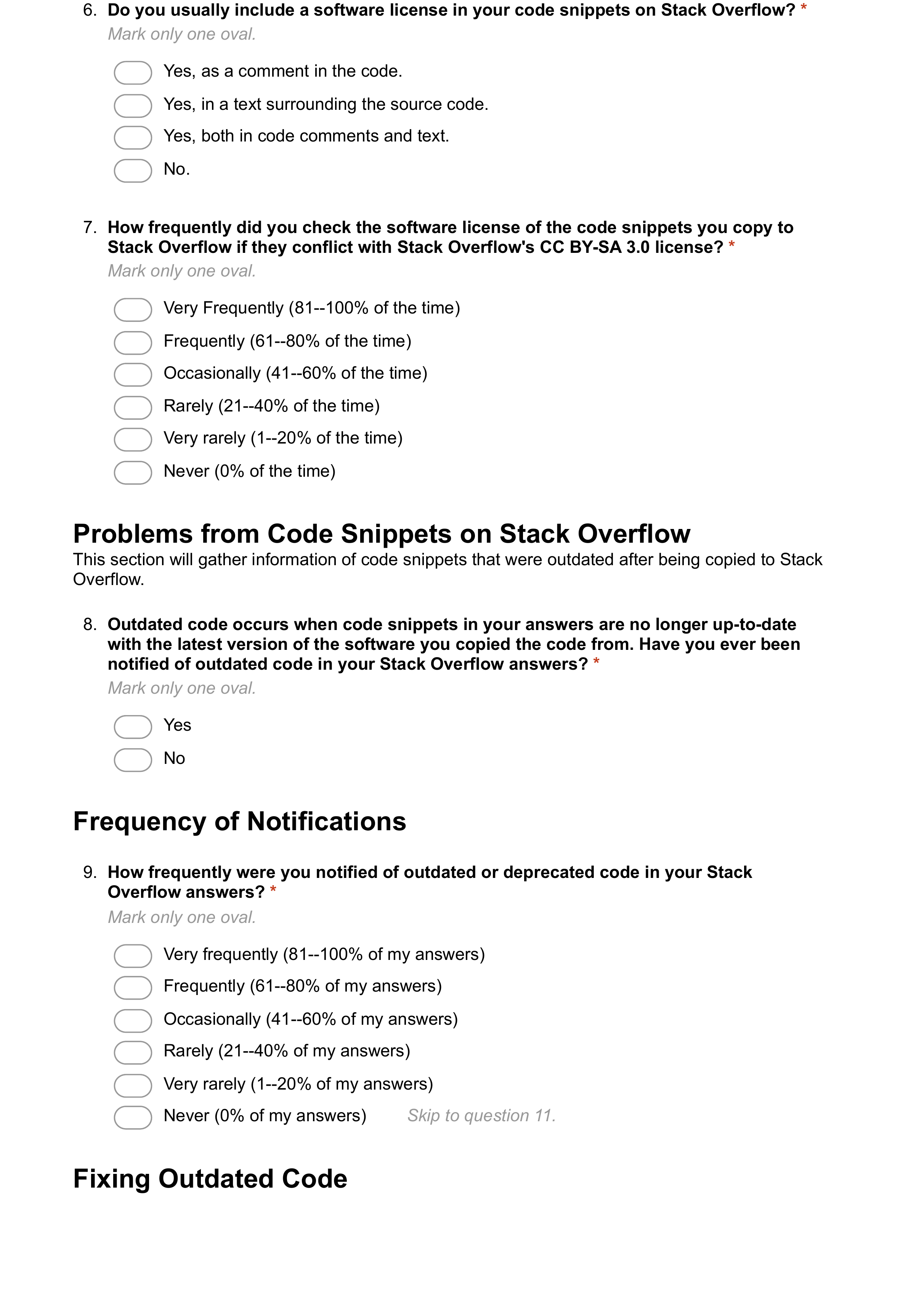}
	\label{fig:answerer-3}
\end{figure}

\begin{figure}[H]
	\centering
	\includegraphics[width=0.9\linewidth]{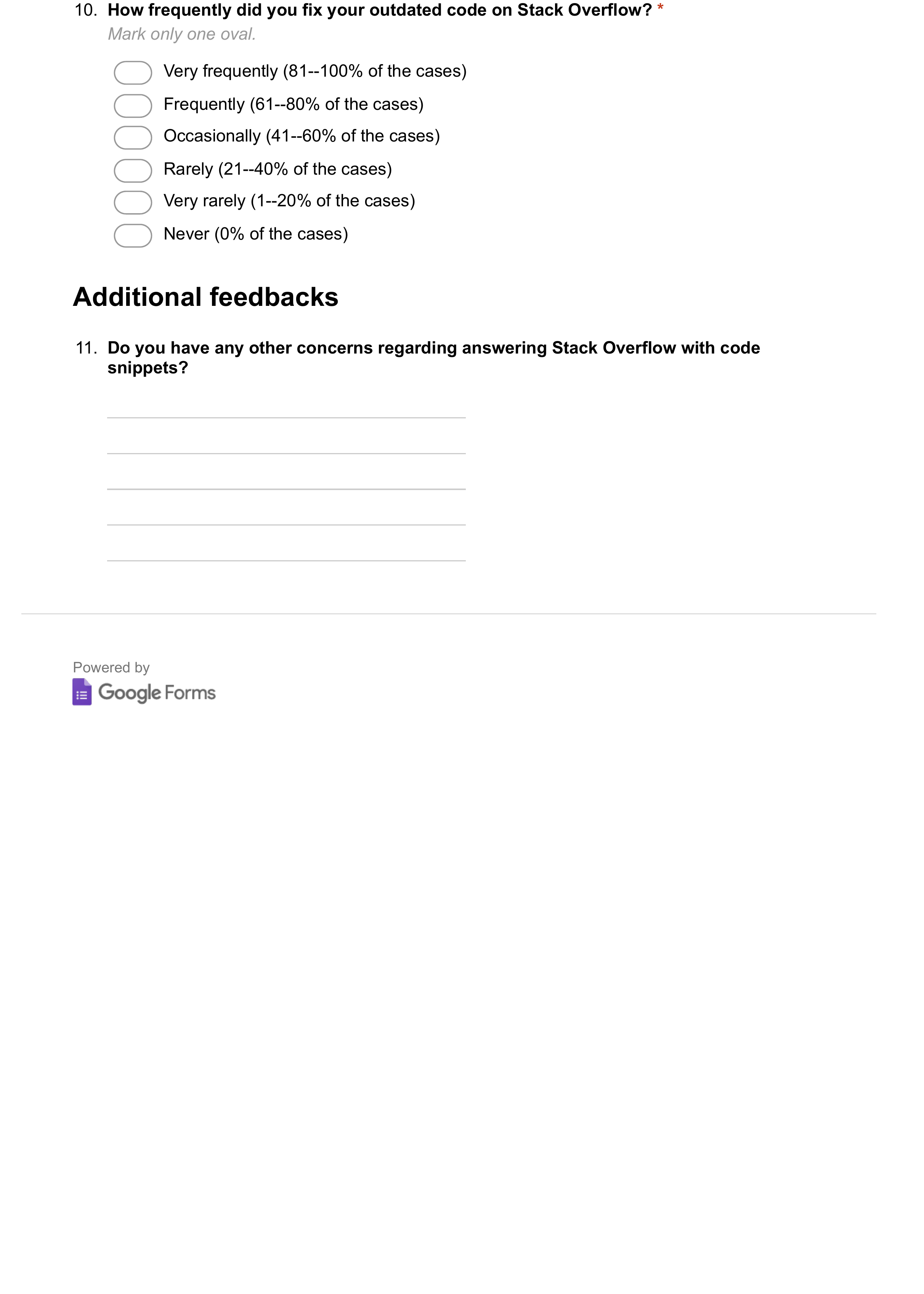}
	\label{fig:answerer-4}
\end{figure}

\section{Stack Overflow Visitor Survey}\label{appendixB}
\begin{figure}[H]
	\centering
	\includegraphics[width=0.85\linewidth]{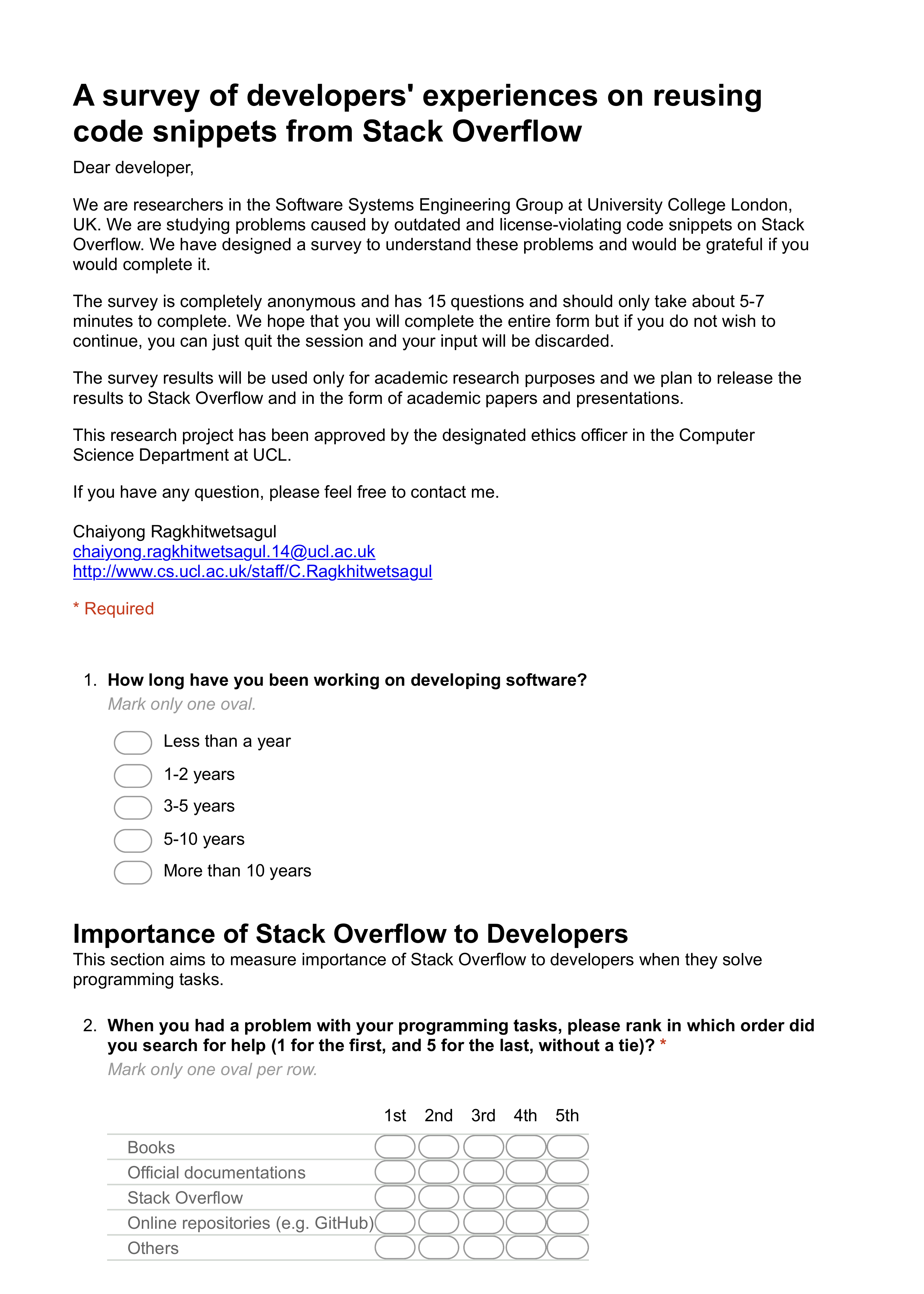}
	\label{fig:visitor-1}
\end{figure}

\begin{figure}[H]
	\centering
	\includegraphics[width=0.9\linewidth]{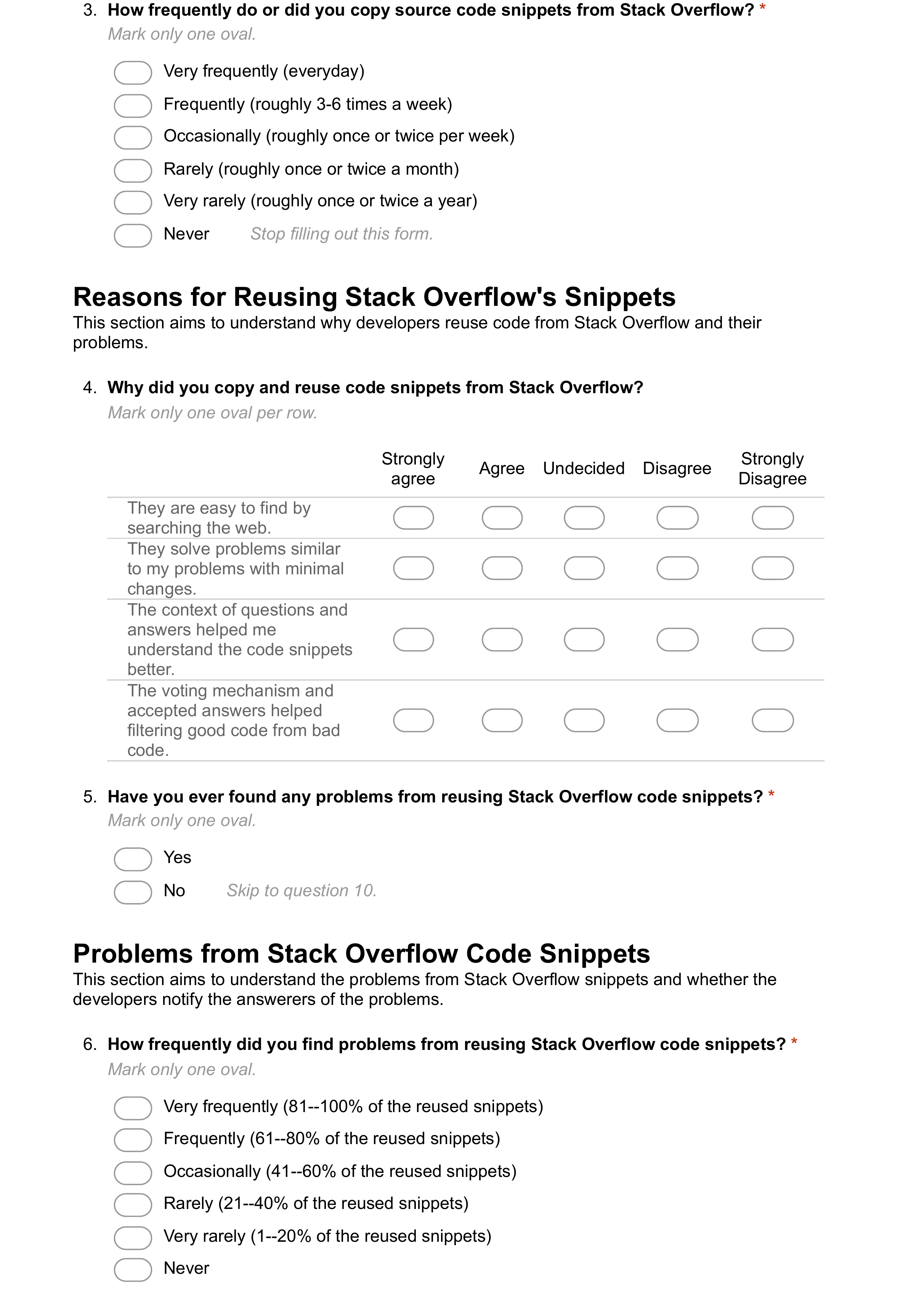}
	\label{fig:visitor-2}
\end{figure}

\begin{figure}[H]
	\centering
	\includegraphics[width=0.9\linewidth]{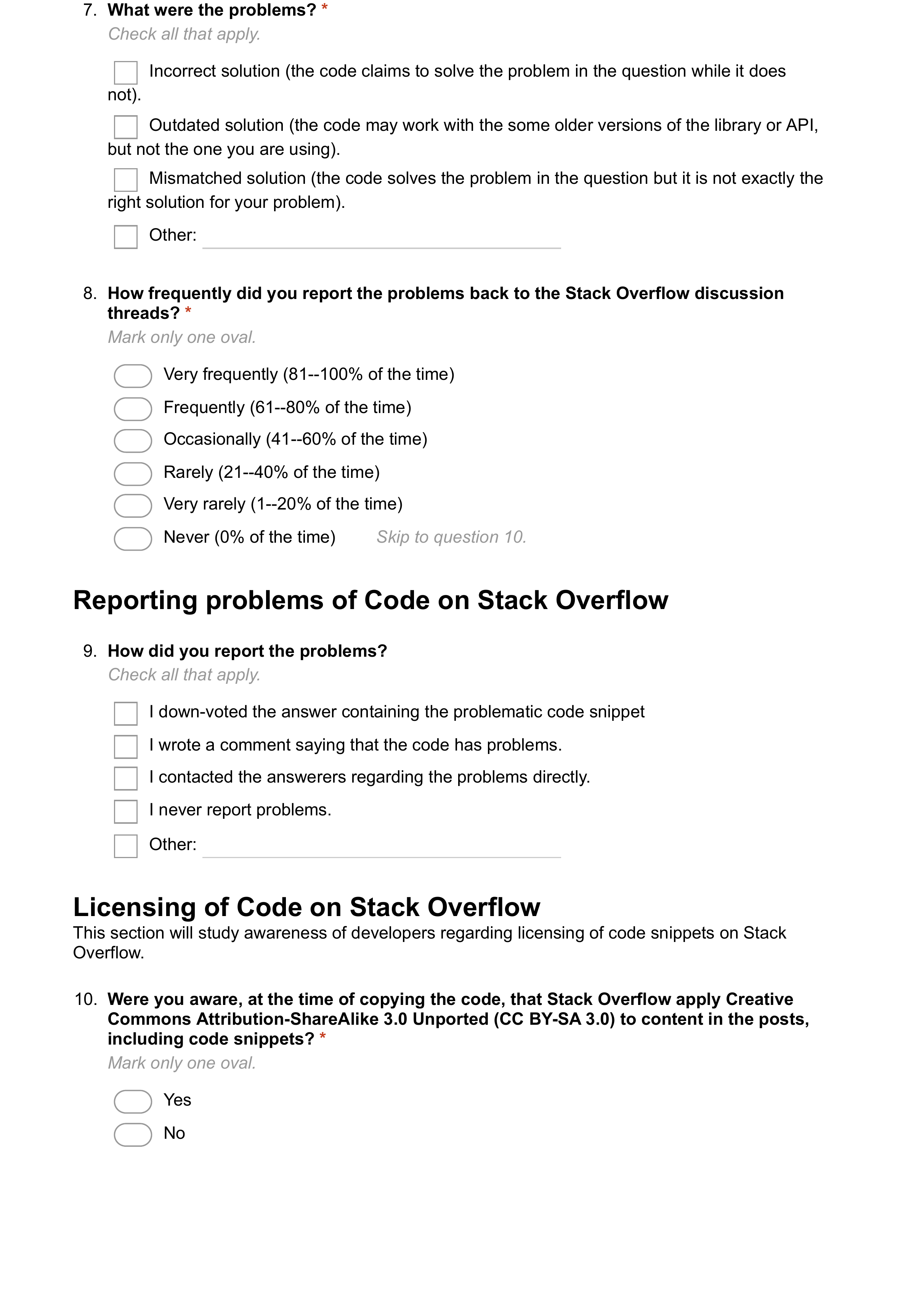}
	\label{fig:visitor-3}
\end{figure}

\begin{figure}[H]
	\centering
	\includegraphics[width=0.9\linewidth]{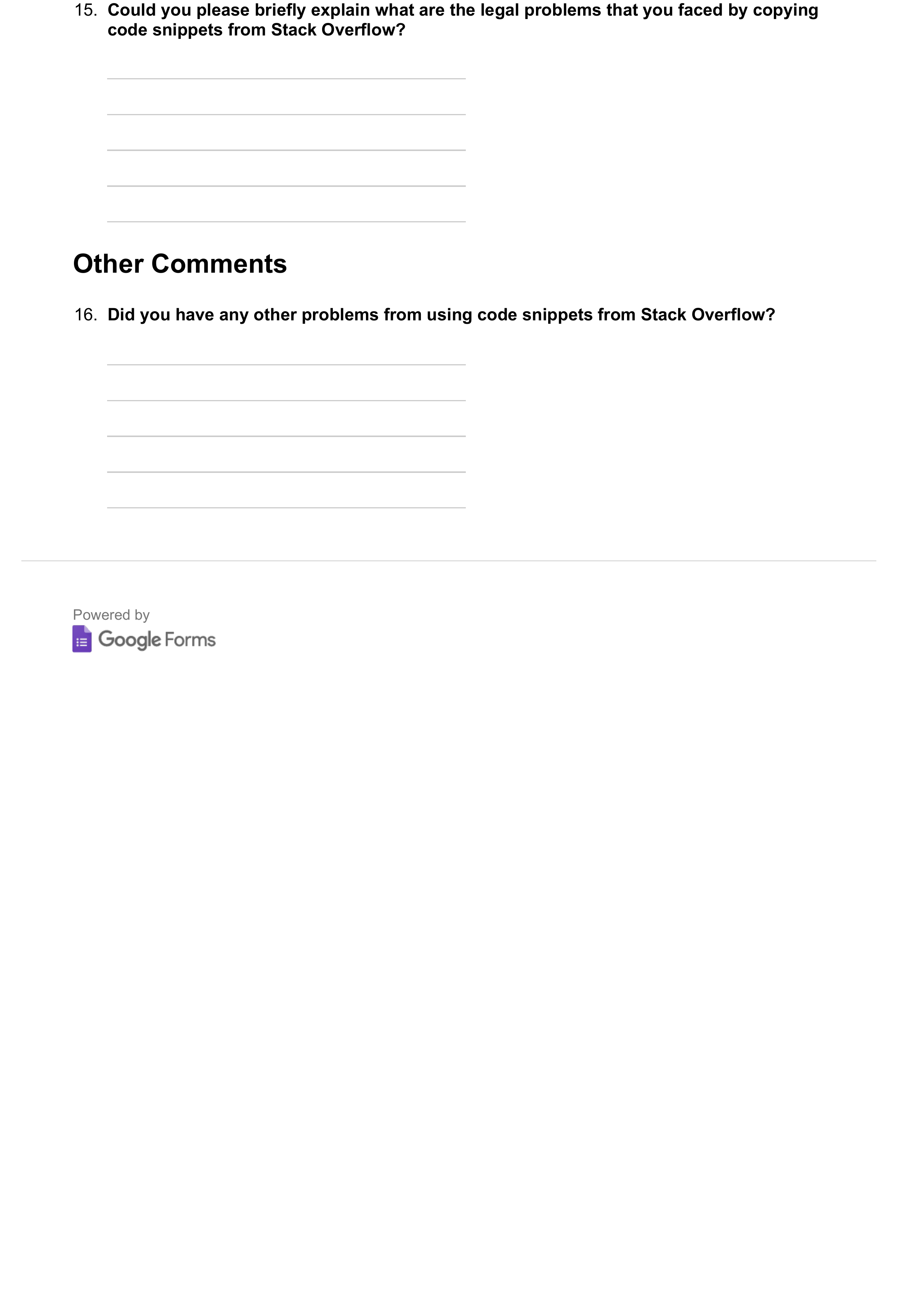}
	\label{fig:visitor-4}
\end{figure}

\section{Open Comments from Stack Overflow Answerers}\label{appendixC}

\begin{enumerate}
	\itemsep1em
	\item Sometimes you have to post code from official documentation, like in case of C\#, code form MSDN is posted in the answer with added explanation.
	\item Snippets on SO are usually for demonstrating a technique and therefore age well. If otherwise, I usually made them a gist, codebin or jsfiddle.
	\item The real issue is less about the amount the code snippets on SO than it is about the staggeringly high number of software ``professionals'' that mindlessly use them without understanding what they're copying, and the only slightly less high number of would-be professionals that post snippets with built-in security issues.
	\item A related topic is beginners who post (at times dangerously) misleading tutorials online on topics they actually know very little about. Think PHP/MySQL tutorials written 10+ years after \texttt{mysql\_*} functions were obsolete, or the recent regex tutorial that got posted the other day on HackerNew (\url{https://news.ycombinator.com/item?id=14846506}). They're also full of toxic code snippets.
	\item Just to say: the reason that I very rarely check the license status is that the code I am posting is almost always my own or adapted from the question, or imported from an open source project that I have worked on and already know the license terms, or from my own company code that I can 100\% say it is OK to post in public because I know our policies.
	\item No. Code snippets are short and small enough that no IP can be in them.
	\item The point of snippets are that they are trivial. I mostly write from scratch or copy-\&-fix a snippet from the question. Most are illustrative or incomplete -- they aren't of any value at all in isolation. They rarely take more than a few minutes to write, and it's usually harder to explain what they're doing in plain English. Where something is large enough to worth the effort of licensing then it's far too big for a snippet. In those cases I create a GitHub project (with a license) and link to it. I'd be wary of increased IP controls -- I doubt there is value they could add to snippets, but they could create significant barriers to contributors, which would hurt the site.
	\item I always try to see what kind of person is asking the question. If it is a student, I don't want to just hand out the answer; they will learn nothing from that. If, on the other hand, it's somebody looking for best practices or a clever trick, I'm not too worried about giving out the solution. In this case, chances are much higher that the person asking the question will go ``Ahh, yes, of course!'' and understand the question, whereas some students are more likely to mindlessly copy-paste the answer.
	\item However, it is also a competitive site, so if your answer requires too much work to incorporate, it won't get accepted or upvoted. As a result, some people---I'm guilty of this myself, I'm sure---will hand out an answer willy-nilly that might solve the problem at hand, but in the long run be a disservice to the person asking.
	\item But I digress: My point is that it's sometimes better to describe a solution rather than just hand off a code snippet. However, since your project seems to pertain to copyright issues, I suppose that's not relevant to you.
	\item I'm not sure it's possible to include a license in your questions/answers. I'm not sure what the legal ramifications would be if you tried it, since you already agreed to S.O.'s terms. This is a very interesting question and I look forward to hearing the results of your research.
	\item I think it's important to realise the code snippets are designed to be very small, useful to illustrate concepts (10 lines or less). When you consider licensing laws of such a small amount of code, while technically may be violating a licence, in practice it would be nearly impossible to enforce such a claim.
	\item SO code snippets are great but there is lot to improve . It's hard to edit and see output in so but website like jsfiddle , jsbin provide nice interface where code editing and output is easy to do.
	\item Outer thing is in so lot of code snippets doesn't work because some users don't add libraries like angular, jquery. I think it's better if we can identify and ask user to auto inject relevant libraries.
	\item Code snippets are usually just a few lines of code so it will be hard to enforce any copyright claims except when it is a method used for something company-specific (such as generating encryption keys). Regardless, since most of the code I write is specifically to answer a given question and having full knowledge of the license system used by Stack Overflow, it is entirely unimportant to concern myself with licensing the code provided. Also, code from MSDN documentation which I sometimes adapt and modify for answers are already in the public domain so it makes no sense re-licensing it.
	\item On the matter of deprecation, I almost entirely use .NET which has got different versions of the framework. Therefore, code deprecation is not often a problem since what is deprecated on one version of the framework may be the only way of solving a given problem on an older version of the framework. I may also have to add that questions I tend to answer are about how to solve general coding problems so they are not usually subject to deprecation.
	\item I think you're forgetting the fact that as a community, we want to share knowledge. Patents, copyright issues and so on -- it's all just annoying. We're there to have fun and to share knowledge with people. 
	\item In the early days, the internet used to be full of free-for-all stuff without any licenses. Because of that, it was a fantastic tool to share knowledge and information on a vast scale. The remnant of this, open source, couldn't have existed without this!
	\item Personally, I believe this ``intellectual property'' drive of the last decade is completely overrated. If you make something \textbf{substantial}, it's fine to be able to claim some ownership -- but on snippets? It's like patenting the stuff you make in your free time in your shed... it doesn't make sense and just adds to the pile of legal bullshit imho.
	\item No. I only put code on there that I have the right to (code I created or have permission to share). Adding the code is not an issue for me.
	\item When I copy code it's usually short enough to be considered ``fair use'' but I am not a lawyer or copyright expert so some guidance from SO would be helpful. I'd also like the ability to flag/review questions that violate these guidelines.
	\item The snippets are all small enough that I reckon they fall under fair use.
	\item I always try and attribute the code I take from other places. I feel pointing back to the origin should be sufficient in terms of giving credit where it's due. Open source is about the sharing of ideas so that others can build on them. Education is a primary use case of open source in my opinion.
	\item My only concern, albeit minor, is that I know people blindly copy my code without even understanding what the code does.
	\item This survey may be inapplicable to me because I never copy code from existing projects.
	\item Stack Overflow did an effort to apply a MIT license to all code snippets, while keeping CC-SA for the text content. Too bad they didn't succeed with it, as it would have solved many issues.
	\item The main problem for me/us is outdated code, esp. as old answers have high Google rank so that is what people see first, then try and fail. Thats why we're moving more and more of those examples to knowledge base and docs and rather link to those.
	\item I've got no issues with code snippets on Stack Overflow, I think they are great. Any one using them should pay attention to details such as the date of the answer etc.
	\item SO's license is not clearly explained when one registers or starts to answer questions.
	\item No, most copied code snippets are so trivial that licensing them would be nearly impossible. It's also mostly modified version, where only some patterns are used.
	\item Lot of the answers are from hobbyist so the quality is poor. Usually they are hacks or workarounds (even MY best answer on SO is a workaround).
	\item I think most example and explanatory snippets don't need a code-specific license. The CC license provides just fine. The examples either aren't copyrightable in the first place, or merely used as starting point (not used exactly as-is, not very different from reading an ``All rights reserved'' education book when learning programming, and ``using'' it in your career every day going forward). In addition, there is also the attitude of authors. Where I might care about attribution for distribution of my answer, the code within my answer is always Public Domain for me, meaning, I would never defend it. (I used to state that on my profile as well, but not in every post.)
	\item Correctness and even syntax are often in doubt if I haven't had time to test the snippet end-to-end under the OP's conditions/environment.
	\item It will be awesome if it becomes simple git repositories like github's gist.
	\item Note that although I was not specifically of SO's licensing terms, I did have an in mind what those terms were likely to be. I have always made sure that there should be no reason that I should not share the code that I included in my replies.
	\item It's an ESSENTIAL part of the site, it would NEVER work without such pieces of code. Also, given the snippets are very small in 99.99\% of cases, legal aspects of this are inherently and pretty much always overlooked by the users.
\end{enumerate}

\end{document}